\documentclass[]{elsarticle}

\usepackage[]{amsmath}
\usepackage{graphicx}
\usepackage{epstopdf}
\usepackage{amssymb}
\usepackage[usenames]{color}
\usepackage[table]{xcolor}
\definecolor{Red}{rgb}{1,0,0}

\usepackage{cases}
\usepackage{verbatim}
\usepackage[left]{lineno}
\usepackage{bm}
\usepackage{upgreek}
\usepackage{makecell} 

\usepackage{textcomp}
\usepackage{multirow}

\newcommand{\dd}{\mbox{\,d}}
\newcommand{\ld}{\ell_{\mathrm{d}}}

\newcommand{\lch}{l_{\mathrm{ch}}}
\newcommand{\olch}{\bar{l}_{\mathrm{ch}}}
\newcommand{\lr}{\ell_{\rho}}
\newcommand{\Gt}{G_{\mathrm{t}}}
\newcommand{\ft}{f_{\mathrm{t}}}
\newcommand{\Gs}{G_{\mathrm{s}}}
\newcommand{\fs}{f_{\mathrm{s}}}
\newcommand{\Kt}{K_{\mathrm{t}}}
\newcommand{\Ks}{K_{\mathrm{s}}}
\newcommand{\oGt}{\bar{G}_{\mathrm{t}}}
\newcommand{\oft}{\bar{f}_{\mathrm{t}}}
\newcommand{\xx}{\bm{x}}
\newcommand{\Ns}{{\ensuremath{N_{\mathrm{sim}}}}}
\newcommand{\seff}{{\ensuremath{s_{\mathrm{eq}}}}}
\newcommand{\eeff}{{\ensuremath{e_{\mathrm{eq}}}}}
\newcommand{\feff}{{\ensuremath{f_{\mathrm{eq}}}}}

\usepackage[
labelfont=bf,
figurename=Fig.,
labelsep=period
]{caption}

\newcommand\Tstrut{\rule{0pt}{2.6ex}}         
\newcommand\Bstrut{\rule[-0.9ex]{0pt}{0pt}}   

\journal{arXiv}

\begin{document}

\begin{frontmatter}
\title{Fracture in random quasibrittle media: I. Discrete mesoscale simulations of load capacity and fracture process zone
}

\author[ism]{Jan Eli\'{a}\v{s}\corref{cor1}}
\cortext[cor1]{Corresponding author}
\ead{jan.elias@vut.cz}
\author[ism]{Miroslav Vo\v{r}echovsk\'{y}}

\address[ism]{Institute of Structural Mechanics, Faculty of Civil Engineering, Brno University of Technology, Veve\v{r}\'{i} 331/95, Brno, 60200, Czech Republic}

\begin{abstract}
Numerical simulations of concrete fracture performed with a~probabilistic mesoscale discrete model are presented. The model represents a~substantial part of material randomness by assigning random locations to the largest aggregates. The remaining part of randomness is introduced by causing material parameters to fluctuate randomly via a~homogeneous random field. An~extensive numerical study performed with the model considers prisms loaded in uniaxial tension with both fixed and rotating platens, and also beams with and without a~notch loaded in three point bending. The results show the nontrivial effect of (i) autocorrelation length and (ii) variance of the random field on the fracture behavior of the model. Statistics of the \emph{peak load} are presented as well as the size and shape of the \emph{fracture process zone} at the moment when the maximum load is attained. Local averaging within the \emph{fracture process zone} and weakest-link are identified as underlying mechanisms explaining the reported results.

The companion paper, Part II~\citep{VorEli19II}, introduces an~analytical model capable of predicting the distribution of the peak load obtained with the probabilistic discrete model via the simple estimation of extremes of a~random field obtained as moving average of local strength.
\end{abstract}


\end{frontmatter}

\section{Introduction}

Concrete fracture behavior is strongly influenced by its internal structure consisting of matrix, mineral aggregates and pores. The material's heterogeneity leads to complex inelastic behavior that still poses a~challenge from the viewpoint of numerical modeling. Over the past few decades, researchers have developed several approaches that allow computer simulations of concrete inelastic behavior. For the sake of this introduction, we make a~distinction between (i) \emph{homogeneous} and \emph{mesoscale} models and (ii) \emph{continuous} and \emph{discrete} models.  

A~\emph{homogeneous} model represents the material at the macroscale without explicit reference to lower scale heterogeneity. The material internal structure (or better the material \emph{characteristic} length) is phenomenologically introduced in the constitutive relation by some internal length parameters. In contrast, \emph{mesoscale} models of concrete directly involve individual heterogeneous units, and the material \emph{characteristic} length arises from the combination of a~constitutive relation at mesoscale and explicitly modeled material structure. It should, however, be noted that the mesoscale models also need to phenomenologically represent fracture processes at an~even lower scale via the transfer of phenomenological information into the constitutive function by user.   

The difference between \emph{continuous} and \emph{discrete} models lies in the assumption made regarding the displacement field. In continuous models, the displacement field is assumed to be continuous and described by some smooth functions with free parameters (degrees of freedom). The governing equations are partial differential equations that are typically discretized in their weak form.  Discrete models assume the displacement field to be established by rigid body motion of discrete units and their governing equations are algebraic (except time derivatives in time-dependent models). Displacement jumps arise between the rigid units. Discrete models are therefore suitable for modeling localized phenomena such as cracks, though their ability to represent elastic behavior is limited. For example, their discrete nature results in restricted Poisson's ratio \citep{BatRot88,AsaAoy-17} or a~boundary layer with different macroscopic elastic properties \citep{Eli17}.  On the other hand, crack representation is difficult for continuous models that otherwise can be used advantageously to describe elasticity and smeared inelastic effects. Finally, there are models that enrich the continuous displacement field by adding discontinuities, and therefore combine the advantages of both approaches. This group includes cohesive zone \citep{CuiLi-19,XueKir19} and extended finite element (XFEM) \citep{MoeDol-99,KumSin-18} models.

In structural engineering, the most frequently used modeling approach is the \emph{homogeneous continuous} description of material with a~continuous displacement field approximated via finite elements. This is because the elastic behavior is of paramount importance and the large dimensions of structural elements do not allow the explicit modelling of the concrete internal structure. In such models, energy dissipation is usually regularized by a~mesh-adjusted constitutive relationship -- the crack band model~\citep{BazOh83}. In order to prevent spurious localization into one finite element layer, there are also localization limiters such as the nonlocal integral \citep{Jir98,HavGra-16} or gradient \citep{PeeBorBreGee:MCFM:98} models, as well as phase field models \citep{YanLi-19}. All these techniques employ some form of internal length parameter as a~user input determining the characteristic length of the missing mesostructure. 

For detailed analysis of the fracture process, on can incorporate the mesoscale structure into continuous models directly. Examples of such \emph{mesoscale continuous} models (often employing cohesive elements) are \citep{ZHOU2019136,ZHANG2019106646,RodOsv-20}. The mesoscale structure can be obtained from X-ray $\mu$CT images of real concrete \citep{HUANG2015340,TraTej-18}. Though these models are probably the most robust, they can only be used to simulate small material volumes due to their excessive computational demands. 

The rigid-body based displacement field used in \emph{discrete} models allows a~significant reduction to be made in the number of degrees of freedom. Discrete models are also advantageous due to their simpler formulation of the constitutive relation in terms of vectors instead of tensors, the discrete nature of the governing equations, the ability to capture the transition from distributed to localized fracture, and naturally oriented cracks. They can be used as \emph{homogeneous} \citep{HwaBol-20,SofVla16,RasMar-18} and applied to real size problems \citep{GedNak12,AmaQia-18}. The discrete homogeneous models are often referred to as Rigid-Body-Spring Network models. Because the discretization does not correspond to the concrete mesostructure, regularization must be performed tu ensure correct energy dissipation. \citet{BerBol06} developed an~efficient regularization technique by adjusting the constitutive relation with respect to discretization density.

Scholars typically employ \emph{discrete} models at the \emph{mesoscale}. The most detailed and also most computationally demanding discrete mesoscale models are lattice models \citep{LukSav-16,PanPra-17,PanMa-18}. The fine discretization into elements is overlaid with realistic mesostructure in order to obtain the information as to whether an~element is found wholly within a~single aggregate, or within matrix, or if it crosses different phases. Its mechanical properties are then modified accordingly. If dynamic effects or  large displacements shall be considered, the discrete element method (DEM) is used \citep{SKARZYNSKI201513,SuchTej-18}. These models with detail mesostructure resolution allows to analyze only small material volumes.
The authors believe that the optimal balance between robustness and computational complexity is to be found in  what are known as lattice-particle models. In such models, ideally rigid discrete blocks represent the stiff mineral aggregates and the displacement jumps account for the deformation of the soft matrix (or the interfacial transition zone) that lies in between \citep{ZubBaz87,IbrahimDelaplace:Discrete:CS:03}. The well-known model of this type is the LDPM developed by Gianluca Cusatis and his collaborators \citep{CusCed07,CusPel-11} and used also by other authors \citep{SheGao-17,FasBol-18}. Though the material volume analyzed by discrete mesoscale models can be relatively large, application to structural elements is still challenging because of the large computational cost. There are several techniques available to reduce the computational complexity such as coarse graining~\citep{LalRez-18}, adaptivity~\citep{Eli16,MikJir17} and low dimensional approximation~\citep{CecZho-18}.

Mesoscale discrete models are now being used to study the size effect on the fracture process zone in notched and unnotched three point bending tests performed on concrete beams \citep{GraGre-12,BenPou-16}. Fracture process zones obtained from randomized discrete mesoscale models have also been studied and compared with those obtained using nonlocal models \citep{GraJir10}, and further used for the calibration of nonlocal models \citep{XENOS201548}. Similarly, damage distributions obtained from 3D lattice analyses with properties dictated by overlaid CT-scans of aggregates have been studied \citep{ZhaGan-19}. The cracking processes obtained using discrete models can be compared with acoustic emission data \citep{GreVer-15} with great success, see also \citep{MurPra-10}.

Despite the accurate information available about material mesostructure, the mesoscale models are incomplete unless they also include spatial variability in material properties that arises during the production process (mixing, drying, etc.) and service life. These spatial fluctuations are often modeled via random fields in both  discrete \citep{GraBaz09,EliVor-15} and continuous \citep{VorSad:08:IJF,GeoSte-14,SyrKor15} models. Even though there are several possible sources of random fluctuations, it is considered here that all of them can be approximately described by a~single homogeneous random field.
As argued in \citep{Vorech:IJSSDogBone:2007,Vor:ComposStru:10}, such a~random field introduces its own characteristic length scale provided in the form of its autocorrelation length, denoted $\lr$. The autocorrelation length and the random field variance become model parameters used to compensate for an~absence of a~more detailed description of random fluctuations in the mesoscale material properties.

We have developed our own in-house simplified version of the LDPM \citep{Eli16,EliVor-15}. It has been enriched by adding spatial variability to the material parameters via a~homogeneous random field. Our numerical model has been thoroughly described in~\citet{EliVor-15}, and the present paper can be understood as a~continuation of that. Therefore, only limited space is devoted here to a~description of the model itself.
The interplay between the \emph{deterministic} internal length, $\ld$, (resulting from the characteristic length of the discrete system) and the \emph{probabilistic} internal length, $\lr$, 
(the autocorrelation length of the random field) is studied with the help of the model. The model is employed to simulate the three point bending of concrete beams with and without a~notch, as well as prisms loaded in uniaxial tension with fixed or rotating platens. The applied random fields are generated with various autocorrelation lengths ranging from $\lr \rightarrow 0$ (independently sampled random variables) up to an~infinitely long autocorrelation length $\lr \rightarrow \infty$, for which the realizations are random constant functions and therefore the whole structure shares the same value in a~single realization. The strengths of the beams, and also the fracture process zones where energy dissipation due to cracking takes place, are statistically evaluated with respect to the effect of the autocorrelation length and the variance of the random field. 

With the help of the data provided here, the companion paper, Part II \citep{VorEli19II}, builds a~novel analytical model capable of delivering the probability distribution function of the peak load.

\section{Probabilistic discrete model}

The model at hand has two fundamental components. The first one of these is the accurate mechanics, which provides stress redistribution, a~transition from distributed to localized cracking and all other features necessary to properly capture the phenomena involved in concrete fracture at the mesoscale level. This also includes the development and propagation of the Fracture Process Zone (FPZ, a~region around a~crack tip that experiences microcracking). The model is a~simplified version of \citep{CusCed07}; the full formulation also accounts for confinement effects and includes more free material parameters. Our model is static; the solution proceeds in loading steps of adaptive length, and iterations are performed to achieve static equilibrium at the end of each step. 

The location and size of the aggregates are randomly generated by the computer based on a~user-supplied sieve curve and the total aggregate volume fraction. The aggregates are simplified to be of spherical shape. The Fuller curve with maximum aggregate diameter 10\,mm, minimum aggregate diameter 4\,mm and total aggregate volume fraction 80\% is used. Starting with the largest aggregates, they are randomly placed within the specimen volume one by one. Each time trial locations are repetitively generated and accepted only after overlapping with previously placed aggregates or specimen boundaries does not occur.

The discrete units of polyhedral shape are obtained by tessellation respecting the layout of the aggregates, each rigid unit contains one aggregate and surrounding matrix. The discrete units are considered ideally rigid with three translational and three rotational degrees of freedom. The rigid body kinematics results in a~displacement jump vector on the contacts between them. A~constitutive function is defined in space determined by contact normal direction, $N$, and two mutually orthogonal tangential directions $M$ and $L$.
\begin{align}  
\left(\begin{array}{c} s_{N} \\   s_{M} \\  s_{L}  \end{array} \right) =  (1-d) E_0 \left(\begin{array}{ccc} 1 & 0 & 0 \\  0 & \alpha & 0 \\ 0 & 0&  \alpha   \end{array} \right) \left(\begin{array}{c} e_{N} \\   e_{M} \\  e_{L} \end{array} \right)
\end{align} 
where $E_0$ and $\alpha$ are two elastic parameters of the model, $\bm{s}$ = stress vector, $\bm{e}$ = strain vector obtained as the displacement jump vector divided by a~contact length $l$ (distance between aggregate centers). $d$ is a~damage parameter calculated based on equivalent stress, $\seff$, and equivalent strain, $\eeff$  
\begin{align}  
d = 1-\frac{\seff}{E_0 \eeff} \quad\quad \text{where} \quad\quad \eeff =\sqrt{e_N^2+\alpha(e_M^2+e_L^2)}
\end{align} 
The equivalent stress reads
\begin{equation}  
\seff = \min\left(\begin{array}{c}(1-d_{\mathrm{prev}})E_0\eeff \\[3mm] \feff \exp\left( \dfrac{K}{\feff}\left\langle \chi -\dfrac{\feff}{E_0}\right\rangle\right) \end{array} \right) \label{eq:seq}
\end{equation} 
The upper expression provides unloading and reloading ($d_{\mathrm{prev}}$ is damage from previous time step) while the bottom one accounts for damage evolution. The angled brackets $\langle \bullet \rangle$ return positive part of $\bullet$, $\feff$ denotes equivalent strength, $K$ is an~initial slope of an~inelastic part of the stress-strain relation and $\chi$ stores history of straining. These variables depend on direction of the straining $\omega$
\begin{equation}  
\tan \omega=\frac{e_N}{\sqrt{\alpha (e_M^2 +e_L^2) }}
\end{equation} 
Pure tension occurs for $\omega=\pi/2$, pure shear for $\omega=0$. 

Let us now focus on predominantly tensile straining direction only since this is the straining mode our simulations experience during inelastic processes. The equivalent strength is defined as
\begin{align}
\feff = \ft \frac{4.52 \sin\omega-\sqrt{20.0704 \sin^2\omega+9 \alpha\cos^2\omega }}{0.04\sin^2\omega-\alpha \cos^2\omega} \label{eq:feq}
\end{align}
with $\ft$ being the tensile strength of the material at mesoscale. The history variable $\chi$ accounting for irreversibility of damage depends on the maximum normal and shear strain throughout the straining history
\begin{align}
\chi = \sqrt{\max\left(e_N^2\right)+\alpha\max\left(e_M^2+e_L^2\right)}
\end{align} 

Finally, $K$ is obtained using $\Kt$ and $\Ks$, the initial slopes of the stress-strain diagram in inelastic regime for pure tension and pure shear,  respectively
\begin{align}
\Kt &= \frac{2E_0\ft^2 l}{2E_0\Gt - \ft^2l} & \Ks &= \frac{18\alpha E_0\ft^2 l}{32\alpha E_0\Gt - 9\ft^2l}
\end{align}
with $\Gt$ being the mesolevel fracture energy in tension. Note that both $\Kt$ and $\Ks$ are dependent on contact length $l$ to preserve energy dissipation (one localized crack is assumed between each two aggregates).
\begin{align}
K = -\Kt\left[ 1-\left( \frac{\omega-\pi/2}{\omega_0-\pi/2}\right)^{n_t}\right] 
\quad\quad \text{where}\quad\quad
n_t = \frac{\ln\left(\Kt/(\Kt-\Ks)\right)}{\ln\left( 1-2\omega_0/\pi\right)}
\end{align}
with $\omega_0$ being the straining direction on the boundary between predominantly compressive and tensile-shear failure.

All of these equations are adopted from \citet{CusCed07} using their recommended relations between remaining inelastic parameters to the governing ones ($\ft$ and $\Gt$): $\fs=3\ft$, $\Gs=16\Gt$, $f_{\mathrm{c}}=16\ft$, $K_{\mathrm{c}}=0.26E$, $\beta=1$, $\mu=0.2$, $n_c=2$. The confinement is omitted, $\lambda_0=0$. Definition of these symbols are not provided here as well as constitutive behavior for predominantly compressive direction, reader is referred to the paper \citep{CusCed07}. Details regarding the simplified constitutive formulation are also published in~\citep{Eli16}. 

The numerical model is (apart from the sieve curve and other geometrical data) controlled by four parameters of the constitutive equation at the contacts:
(i) the elastic modulus, $E_0$
(ii) the tangential/normal stiffness ratio $\alpha$, (both of which control the elastic behavior); (iii) the mesolevel tensile strength, $\ft$, and (iv) the mesolevel fracture energy in tension, $\Gt$, (both of which control the inelastic behavior).

The basic version of the model itself provides a~random response due to its random mesostructure: there is randomness in the positions and sizes of the discrete bodies (and thus in the dimensions and orientation of the contacts). However, we will refer to the model without random fluctuations of material parameters as the \emph{deterministic} model hereinafter, and the mean value and standard deviation of the peak load provided by the deterministic model will be denoted $\mu_{\mathrm{d}}$ and $\delta_{\mathrm{d}}$, respectively.

The second fundamental component is an~additional random spatial fluctuation of the material parameters. The combination of the deterministic model with this randomization will be called the \emph{probabilistic} model, and the corresponding mean and standard deviation of the peak load will be denoted $\mu_{\mathrm{p}}$ and $\delta_{\mathrm{p}}$, respectively.

In the probabilistic model, the two material parameters governing fracture behavior ($\ft$ and $\Gt$) are considered to vary randomly in space. In particular, one homogeneous dimensionless random field $h(\xx)$ is assumed to control the values of both parameters at any position $\xx$ of the contacts. The cumulative distribution function $F(h)$ of the random field is assumed to be Gaussian with a~left Weibullian tail \citep{BazPan:Activation:07:JMPS,LeBaz-11}. The mean value is one, $\mu_h=1$, and the standard deviation, $\delta_h$, is a~free parameter to be identified from experiments.
The random field is discretized to obtain a~set of random variables that represent the field at the centroids of the contact faces. The correlations among all pairs of these variables are dictated by an~isotropic correlation structure with a~separable squared exponential (correlation) function governed by the second free parameter, the autocorrelation length $\lr$ 
A~detailed description of the probabilistic model and an~efficient method for generating the random field samples can be found in \citep{EliVor-15}.

The additional randomness due to the spatial variability of $h(\xx)$ is incorporated into the constitutive relation via the modification of the fracture parameters, $\ft$ and $\Gt$, that would be associated with the contacts in the deterministic version of the model. Since the inelastic shear parameters ($\fs$ and $\Gs$) are just multiplications of the tensile parameters, both shear and tensile fracture behavior actually vary randomly at the same time. Two alternative methods of varying the fracture energies are explored here. In \emph{Alternative I}, both the parameters are considered linearly dependent on the random field value
\begin{align}
    \ft(\xx)&=\oft  h(\xx) & \Gt(\xx)&=\oGt  h(\xx)
    \label{eq:scaling1}
\end{align}
where $\oft$ and $\oGt$ stands for the deterministic model parameters. A~noticeable feature of this alternative is that the mesoscale Irwin characteristic length at mesoscale, which is constant in the deterministic model, becomes randomly varying in space
\begin{align}
    \lch (\xx) = \frac{E  \Gt(\xx)}{ \left[\ft(\xx)\right]^2}
    =
    \frac{E  \oGt  h(\xx)}{ \left[\oft  h(\xx)\right]^2}
    =
    \frac{\olch}{h(\xx)}
    \label{eq:ldet1}
\end{align}

\begin{figure}[tb]
\centering
\includegraphics[width=14cm]{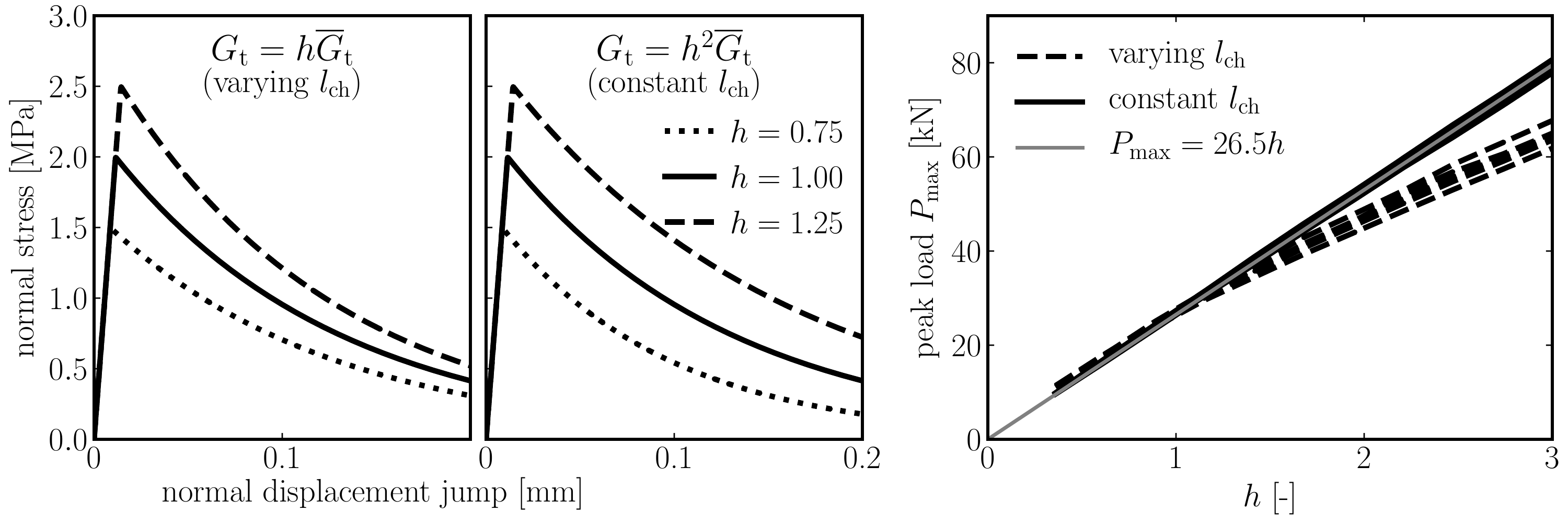}
\caption{Left: the constitutive function in monotonic normal loading for both described alternative methods of fracture energy randomization; right: peak load measured on several models with different mesostructure and with a~constant $h$ value for all contacts.\label{fig:scaling}}
\end{figure}

Even though the alternative with random $\lch (\xx)$ might be applied in general,  we use the second alternative, in which the characteristic length is kept constant (as originally suggested in~\citep{VorSad:08:IJF}).
\begin{align}
    \ft(\xx)
    &=
    \oft h(\xx) & \Gt(\xx)
    &=
    \oGt \left[h(\xx)\right]^2
    \label{eq:scaling2}
\end{align}
In this \emph{Alternative II}, the mesoscale characteristic length at any contact equals the ``deterministic'' characteristic length
\begin{align}
    \lch(\xx)
    =
    \frac{E  \Gt(\xx)}{ \left[\ft(\xx)\right]^2}
    =
    \frac{E  \oGt \left[h(\xx)\right]^2}{ \oft^2 \left[h(\xx)\right]^2}
    =
    \frac{E  \oGt}{ \oft^2}
    =
    \olch
    =
    \mathrm{const.}
    \label{eq:ldet2}
\end{align}
Due to the squared value of $h(\xx)$ for the random field of fracture energy in Eq.~\eqref{eq:scaling2}, the fracture energies become, on average, slightly higher than the material parameter $\oGt$ by a~factor $\mu_{h^2} = \delta_h^2+\mu_h^2=\delta_h^2+1$. In this paper, it is assumed that $\delta_h$ varies in range from 0.035 to 0.28 and that therefore the fracture energies in Alternative II are higher by 0.1225\% to 7.84\% on average compared to those of the deterministic model (or Alternative I). This is considered an~insignificant difference that is well compensated by the advantages of having a~constant mesoscale characteristic length, i.e. one which is independent of random field $h(x)$.

Both alternatives are depicted in Fig.~\ref{fig:scaling}, where the response of a~single contact loaded monotonically in the normal direction is plotted for different values of $h$. In Alternative I, the \emph{deterministic} internal length decreases with an~increase in $h$ and the contact becomes more brittle, see Fig.~\ref{fig:scaling} left.

Imagine that a~constant value of $h$ is used to modify the parameters of the whole structure. Such a~situation actually corresponds to one extreme case described below -- the case of an~infinitely large autocorrelation length $\lr\to\infty$. In such a~case, Alternative II leads to the linear dependence of the structural strength on $h$. The sequence of events (such as damage evolution, etc.) in the model is exactly the same for any positive value of $h$. In contrast, the first alternative produces a~nonlinear dependence on $h$ as the redistribution of stress in the mechanical system is affected by $h$. This is demonstrated in Fig.~\ref{fig:scaling} right, including a~linear approximation line for $\mu_{\mathrm{d}}=26.5h\,$kN. This fundamental property of the second alternative, caused by constant mesoscale characteristic length, will later allow us to separate the randomness from the aggregate positions and randomness from the random field for one specific case of infinite autocorrelation length. This is the only reason while we are using alternative II. Apart from this advantage, both alternatives can be equally well applied. The same version of fracture energy probabilistic scaling, which keeps $\lch$ constant in space, has recently been used for the dynamic simulation of ceramics~\citep{LeEli-18}. 

In order to show different randomization techniques and their effect on the results, we included \ref{sec:C}, where short study considering different model variants as well as randomization according to alternative I is presented.

\section{Identification of material parameters \label{sec:identification}}
The material parameters of the mechanical model and its probabilistic extension are partly identified and partly fabricated. The identification is based on an~experimental campaign described in \citep{GreRoj-13}, which was performed on concrete with a~10\,mm maximal aggregate diameter. Beams of different size and notch depth were tested in three point bending. The beams depths were 50, 100, 200 and 400\,mm, relative notch depths were 0.5, 0.2 and 0. The number of tested specimens for each case varied from 2 to  5; usually, 3 specimens were tested.

\begin{figure}[tb!]
	\centering
	\includegraphics[width=\textwidth]{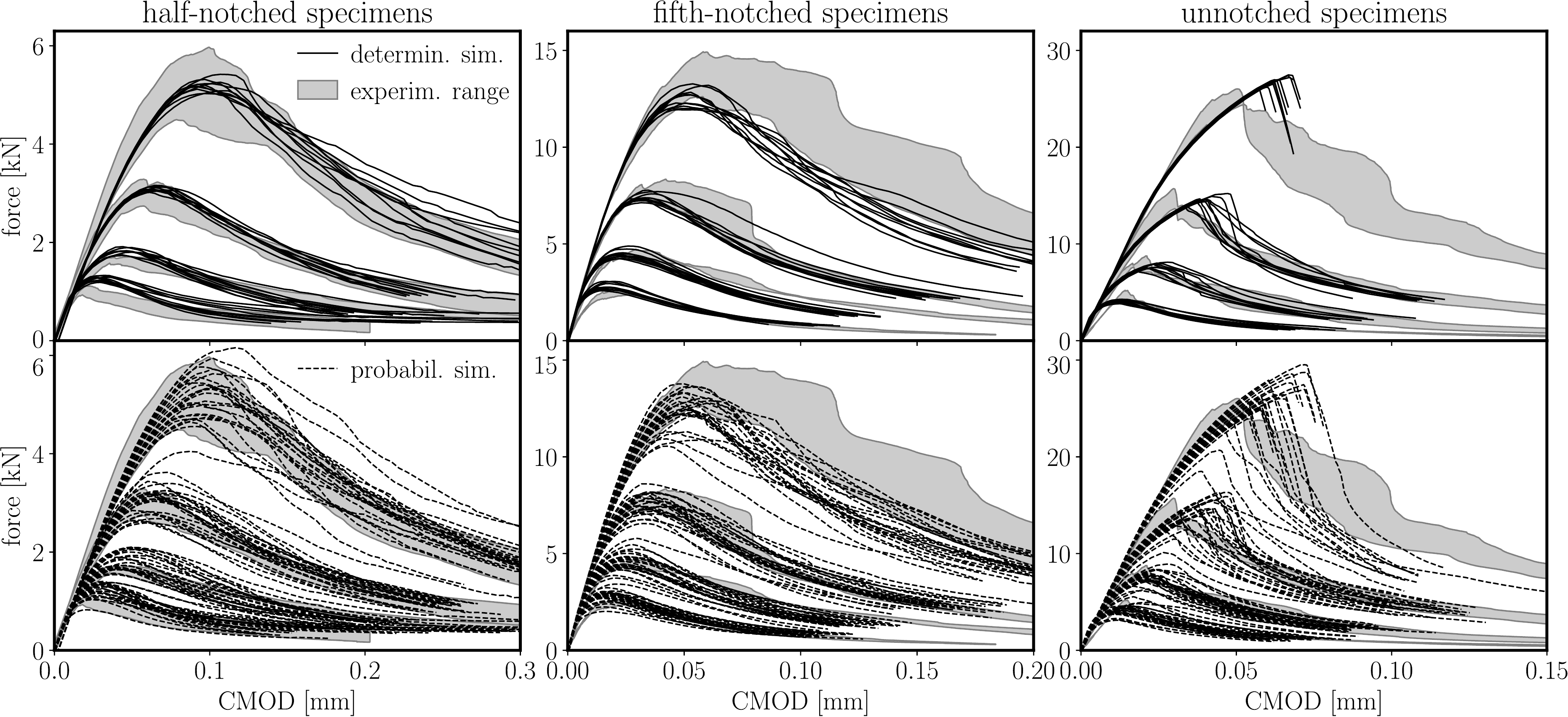}
	\caption{Comparison of results of the \emph{determinitic} and \emph{probabilistic} models with the experimental data from~\citep{GreRoj-13}. The top row shows 10 simulations with the deterministic model with random aggregate layouts, the bottom row shows 24 simulations with the probabilistic model with random aggregate layouts and different random field samples. The left, central and right columns present results of the half-notched, fifth-notched and unnocthed specimens, respectively; each graph combines data for four different beam sizes. \label{fig:comp_experiments}}
\end{figure}

The $\alpha$ parameter is set to 0.24 which provides Poisson's ratio of about 0.21. The second elastic parameter, the mesoscopic elastic modulus $E_0$, is determined by a~trial end error method from the initial elastic part of the experimental response of large beams to be 27\,GPa. The mean values of both inelastic parameters, $\oft$ and $\oGt$, are found in automatic optimization using Nelder--Mead (downhill simplex) method. The objective function subjected to optimization is sum of squared relative errors in (i) maximum peak load and (ii) area under load--CMOD curve for three largest sizes (100, 200 and 400\,mm) and both notched cases. The unnotched case is omitted as we expected a~large effect of randomness in the experimental measurements. The smallest size is omitted because it has too low number of aggregates in the ligament and the model becomes too coarse. Besides, simulations of specimens too small with respect to their sieve curve are largely affected by the boundary layer \citep{Eli17}. The mesoscopic fracture energy and tensile strength were identified to be $\oGt=25$\,J/m$^2$ and $\oft=2$\,MPa, respectively. Comparison of the deterministic model results and the experimental data is shown in Fig.~\ref{fig:comp_experiments}, top row. All sizes and notch depths are shown, not only those involved in the optimization process. Results from 10 different random aggregate layouts are plotted, however only one layout was used during calibration. Considering that only two model parameters were subjected to optimization, we see the results as confirmation of model robustness and its validation.

\begin{table}

\centering
\caption{Differences between three experimentally and numerically obtained descriptors. The first column for each descriptor shows the values obtained in experiments \citep{GreRoj-13}, the second and third columns show relative differences of numerical results of the deterministic and probabilistic models, respectively. \label{tab:identification}}
\tabcolsep 4.5pt
\begin{tabular}{ccc|ccc|ccc|ccc}
\Xhline{4\arrayrulewidth}
 \multirow{2}{*}{$\alpha_0$} & $D$ & num. & \multicolumn{3}{c|}{mean $P_{\max}$} & \multicolumn{3}{c|}{st. dev. $P_{\max}$} & \multicolumn{3}{c}{mean CMOD$_{\max}$} \\ 
 & mm & exp. & kN & \% & \% & kN & \% & \%  & \textmu m & \% & \% \\\hline\hline
\multirow{4}{*}{0.5}	&	400	&	3	&	5.2	&	0.0	&	-0.6	&	0.62	&	-82.5	&	-20.9	&	90.6	&	8.5	&	6.4	\\
&	200	&	5	&	3.0	&	3.7	&	2.2	&	0.23	&	-81.8	&	16.2	&	57.9	&	11.4	&	9.8	\\
&	100	&	2	&	1.7	&	10.3	&	9.1	&	0.05	&	5.7	&	288.8	&	35.7	&	28.0	&	24.0	\\
&	50	&	3	&	1.0	&	29.8	&	25.8	&	0.09	&	-49.0	&	45.9	&	19.3	&	51.3	&	47.7	\\\hline
\multirow{4}{*}{0.2}	&	400	&	3	&	14.0	&	-10.7	&	-11.3	&	1.02	&	-55.3	&	-13.3	&	58.1	&	-12.5	&	-11.1	\\
&	200	&	3	&	7.8	&	-4.4	&	-3.6	&	0.41	&	-52.9	&	49.4	&	37.6	&	-12.4	&	-9.4	\\
&	100	&	3	&	4.5	&	-0.7	&	0.1	&	0.16	&	21.5	&	183.4	&	29.2	&	-19.9	&	-18.0	\\
&	50	&	2	&	2.6	&	9.5	&	8.0	&	0.15	&	10.6	&	79.3	&	21.3	&	-20.0	&	-21.2	\\\hline
\multirow{4}{*}{0}	&	400	&	2	&	25.2	&	6.6	&	1.7	&	0.80	&	-43.2	&	201.4	&	51.0	&	23.4	&	15.9	\\
&	200	&	3	&	14.5	&	0.2	&	-2.2	&	0.88	&	-54.0	&	70.3	&	30.2	&	29.8	&	25.0	\\
&	100	&	3	&	8.1	&	-5.5	&	-8.2	&	0.50	&	-42.5	&	50.8	&	21.2	&	6.9	&	1.5	\\
&	50	&	2	&	4.8	&	-13.7	&	-13.2	&	0.45	&	-67.3	&	2.5	&	14.9	&	-17.3	&	-9.0	\\
\Xhline{4\arrayrulewidth}
\end{tabular}
\end{table}

The probabilistic part of the model has four independent parameters to describe the probability density function of $h$, along with one parameter related to spatial correlation: the autocorrelation length, $\lr$. The mean value of $h$ must be equal to 1 according to the randomization procedure. The Weibull modulus is assumed to be 24 and the grafting probability $P_{\mathrm{gr}}=10^{-3}$ is chosen (meaning that the values
of $h$ that are Weibull-distributed occupy only the far left tail up to a~probability 10$^{-3}$). We remark that the Weibullian
tail has only minor impact as the corresponding values are very rarely featured in samples of random fields.
The standard deviation of $h$ is identified from the experimental results on half-notched specimens with depth 200\,mm, because 5 samples (the largest number) were tested in this group. A~minimization of the difference between the peak load standard deviations from experiments and 24 simulations yields $\delta_h=0.14$. The autocorrelation length is estimated based on the peak loads of the three largest unnotched specimens. Based on analysis with 24 random simulations, the autocorrelation length is found to be approximately $\lr=100$\,mm. Fig.~\ref{fig:comp_experiments}, bottom row, presents comparison of the experimentally measured data with results of 24 random samples of the calibrated random model for all sizes and notch depths. 

Tab. \ref{tab:identification} provides relative differences between experiments and simulations using the following descriptors: (i) the average peak load, (ii) the standard deviation of the peak load and (iii) the average CMOD at the peak load (for experiments it is the CMOD at the peak of the average load-CMOD curve). The difference in the average peak load is typically up to 15\%, often less then 5\%.  Only for the smallest half-notched specimens the difference increases up to 30\% because the ligament is extremely small (25\,mm) 
compared to the maximum aggregate size (10\,mm). The comparison of the standard deviations of the peak loads is in most of the cases impaired
by very low numbers of specimens tested (column 3 in Tab.~\ref{tab:identification}). Differences up to 30\% appear for the average CMOD$_{\max}$, where both the deterministic and probabilistic models exceed experimental data for half-notched and unnotched specimens, but underestimate them for fifth-notched specimens. An~exception is again the smallest half-notched specimen with large differences due to poor discretization of the ligament.

The differences in descriptors obtained with the \emph{probabilistic} model are typically lower compared to those obtained with the \emph{deterministic} model. Application of the artificial random field compensating for all missing sources of randomness in the model is therefore justified; see also additional arguments presented in Section~2 of the companion paper, Part II~\citep{VorEli19II}.

We now use the calibrated model to find out what happens when the autocorrelation length, $\lr$, and standard deviation, $\delta_h$, of the random field vary. How do these variations affect the maximum load and crack pattern? Is the effect of these two free parameters different for different boundary conditions?

\begin{figure}[tb!]
\centering
\includegraphics[width=\textwidth]{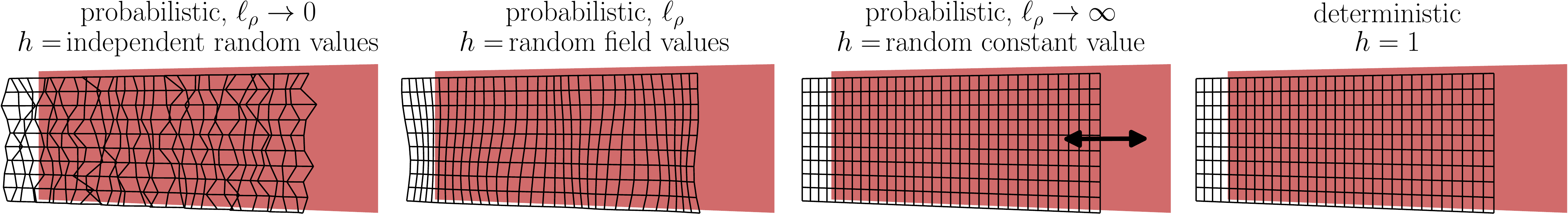}
\caption{2D illustration of the $h(\xx)$ parameter in the \emph{deterministic} model and the \emph{probabilistic} models with autocorrelation lengths $\lr\in\left\{0,25,\infty\right\}$\,mm. \label{fig:RF}}
\end{figure}

\section{Simulated beams and load capacity
\label{sec:simulations}}

We study the interplay between mechanical and probabilistic components by performing sets of simulations of beam responses with various realizations of the random fields. The same actually applies to the deterministic model as the arrangements of the aggregates imply variance in the response characteristics. To estimate the mean value and standard deviation of the load capacity (peak load) and to process the crack patterns, $\Ns=100$ realizations of every model variant are performed. The realizations differ in their random mesolevel arrangements, and additionally in terms of the random field realizations (if used).

The major focus is now on the interaction between two governing material lengths, the autocorrelation length $\lr$ that is varied in the numerical study, and the macroscopic characteristic length $\ld$ established naturally in the mesoscale model by the generated model mesostructure and by the application of constitutive relations. The characteristic length is difficult to control explicitly, and therefore we keep the sets of generated mesostructures and constitutive relationships unchanged in most of the studies. The autocorrelation length parameter is easy to modify. Several values of $\lr$ ranging from $\lr \to 0$ (which represents a~situation with mutually independent random field values) to
$\lr \to \infty$ (where random values are identical throughout the whole volume for each sample) are considered. For intermediate $\lr$, a~sample of a~random field with a~corresponding autocorrelation length and a~square exponential covariance function is generated and applied, see Fig.~\ref{fig:RF}.

The second probabilistic parameter subjected to change is the standard deviation (since $\mu_h=1$, it is equivalent to the coefficient of variation) of the random field $h(\xx)$. It changes the intensity with which randomness is applied. The basic value approximately identified from experimental data is $\delta_h=0.14$, though three more additional values are considered: $\delta_h=0.035$,  $\delta_h=0.07$ and  $\delta_h=0.28$.

In total, $\Ns=100$ different mesostructure arrangements are used repetitively for each model version. This means that the $i$th simulation of the deterministic model shares the same arrangement with the $i$th simulation of all probabilistic models irrespective of autocorrelation length or variance. In the case of a~random field with $0<\lr<\infty$, $\Ns=100$ different random field realizations are generated for each autocorrelation length using $\delta_h=0.14$. These realizations are then rescaled to different standard deviations
\begin{align}
\left[h(\xx)\right]_{\delta_h} = 1+\frac{\left[h(\xx)\right]_{0.14}-1}{0.14}\delta_h \label{eq:scalingGrafted}
\end{align}
where $\left[h(\xx)\right]_{\delta_h}$ denotes a~realization of a~random field with standard deviation $\delta_h$. The transformation preserves the Gaussian part, but changes the grafting probability and distorts the Weibullian tail. When $\delta_h>0.14$, the Weibullian tail protrudes into negative values and must be trimmed to ensure positive values of $h(\xx)$. As noted earlier, the probability of reaching the Weibullian tail is extremely low and therefore the results are not notably affected by such incorrect transformation. In the case of zero autocorrelation length, the random field values $h(\xx)$ are drawn for each contact randomly and independently from the grafted distribution with $\delta_h=0.14$ and rescaled according to Eq.~\ref{eq:scalingGrafted} if needed. Finally, for an~infinite autocorrelation length, all the contacts in one simulation share the same (possibly rescaled) value of $h$  generated using stratified sampling. The $i$th simulation has a~value of $h$ obtained by the inverse probability density function of the distribution $F_h^{-1}$: $h_i=F_h^{-1}((i-0.5)/\Ns)$. 

\begin{figure}[tb!]
\centering
\begin{tabular}{cc}
\includegraphics[width=15cm]{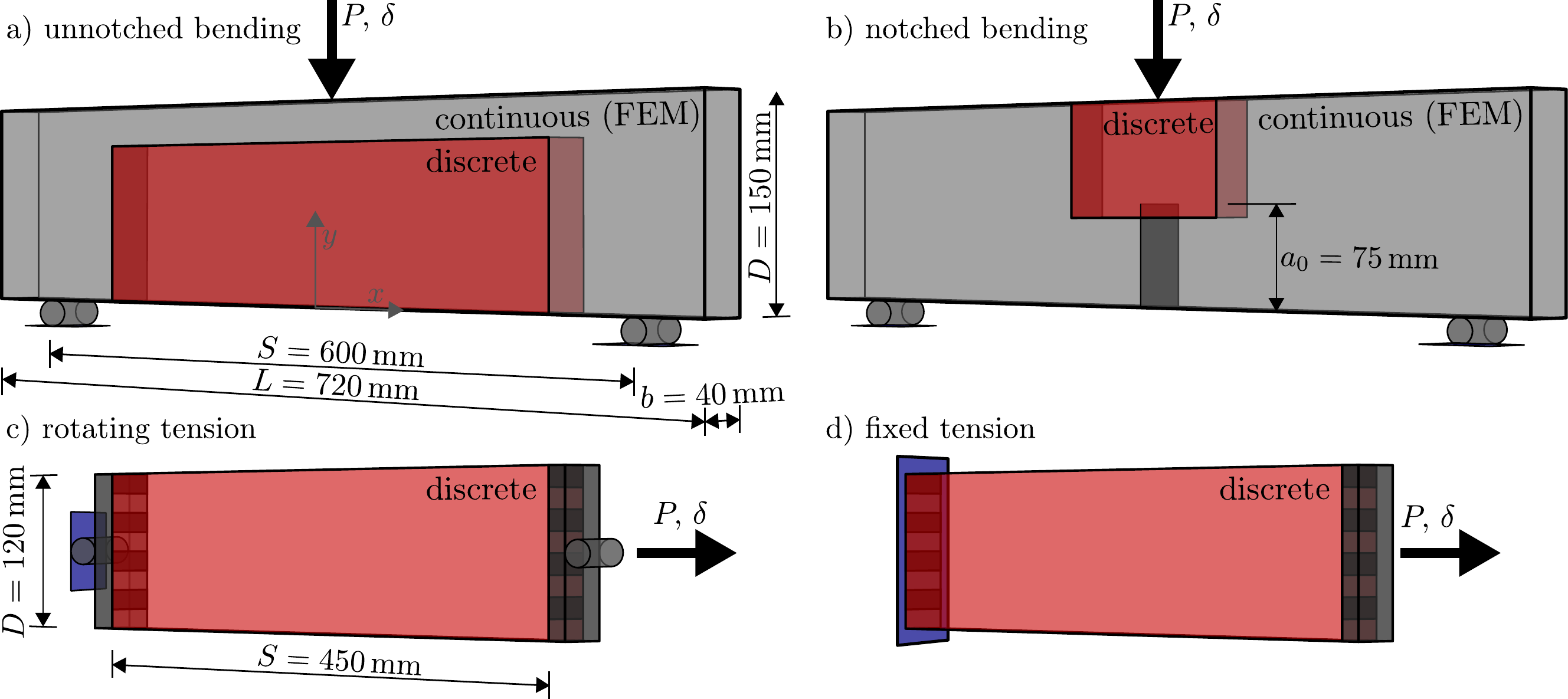}
\end{tabular}
\caption{Virtual specimens geometry and coordinate system - a) unnotched and b) notched beams loaded in three point bending; prisms loaded in uniaxial tension assuming c) rotating and d) fixed platens at both ends. \label{fig:scheme}}
\end{figure}

\begin{figure}[tb!]
	\centering
	\begin{tabular}{cc}
		\includegraphics[width=16cm]{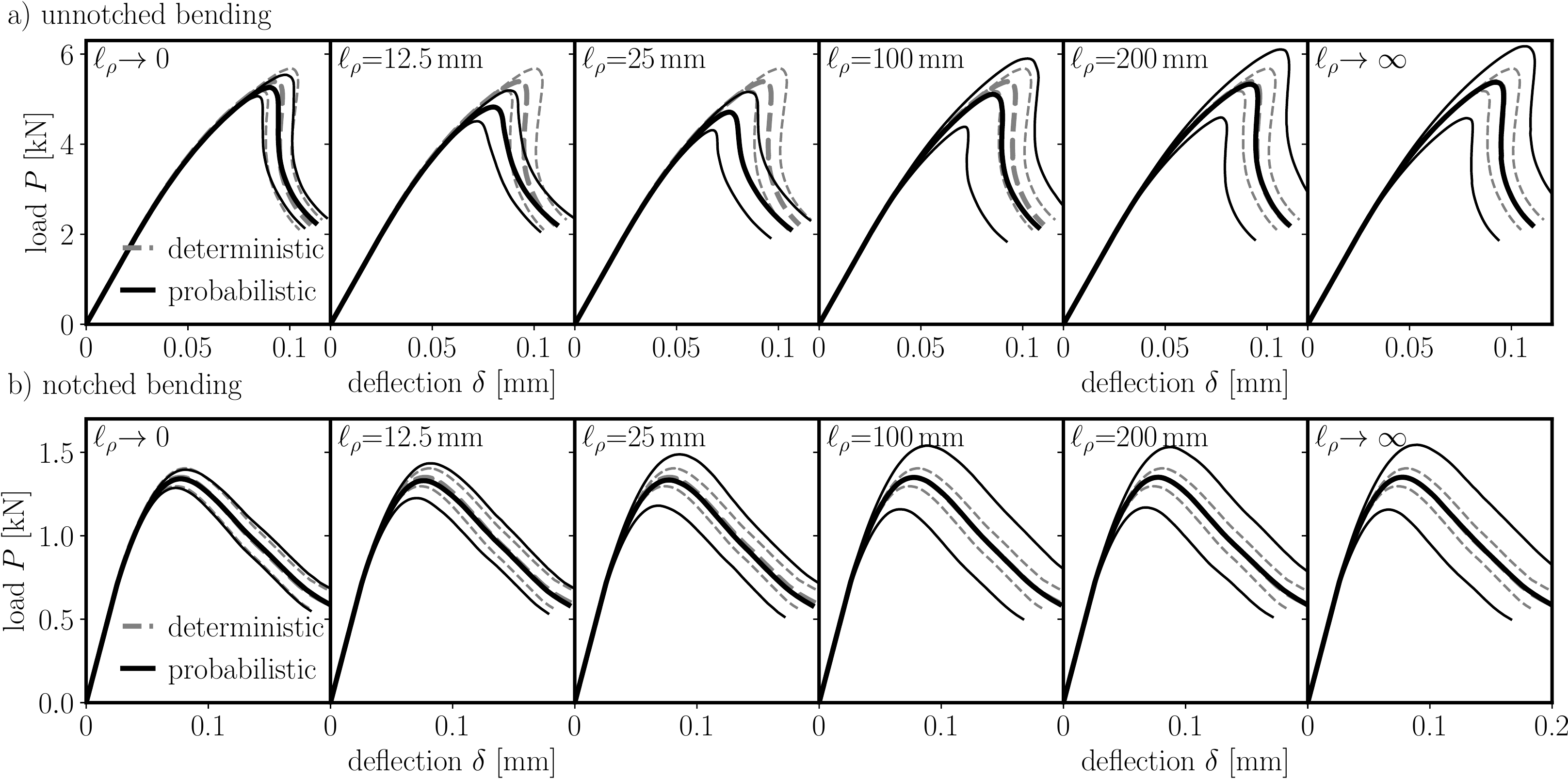}
	\end{tabular}
	\caption{Average load deflection curves and bands of one standard deviation width calculated from 100 simulations of deterministic and probabilistic models. Only results from the probabilistic model using $\delta_h=0.14$ are presented, other standard deviations of the random field yield similar curves. \label{fig:LD_curves}}
\end{figure}

\begin{figure*}
\centering
\includegraphics[width=\textwidth]{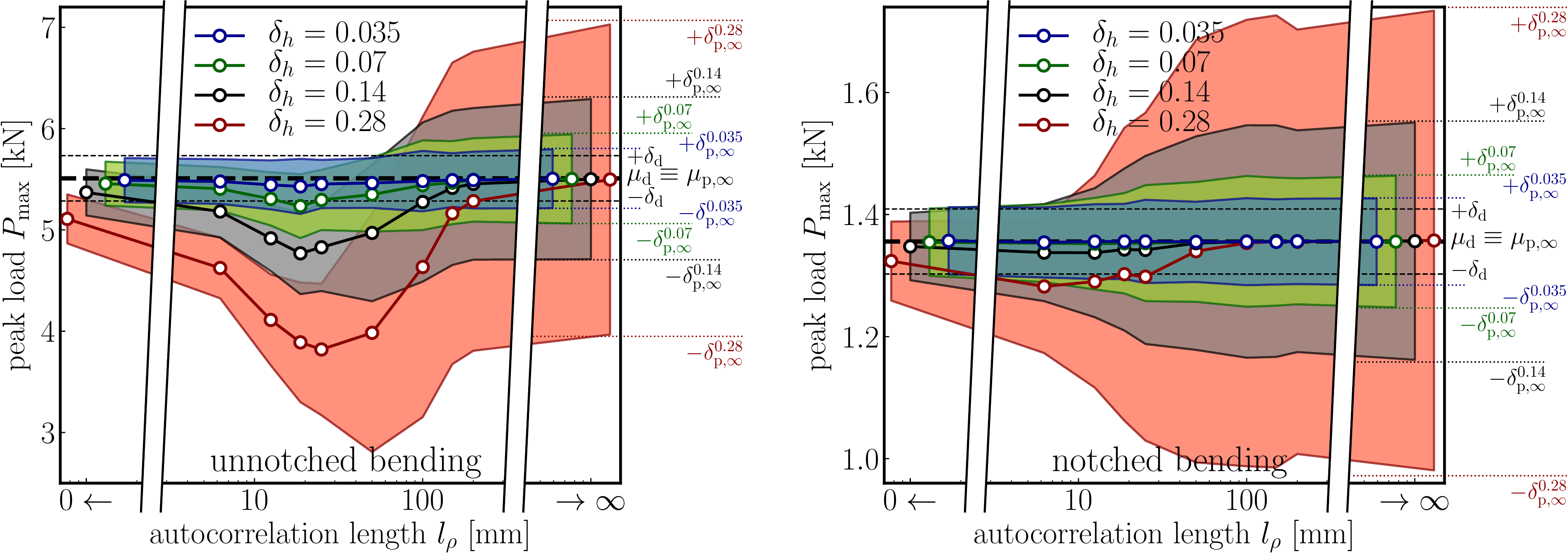}
\caption{
    The mean value and standard deviation of the maximum load computed on unnotched (left) and notched (right) beams loaded in three point bending using the \emph{deterministic} ($\mu_{\mathrm{d}}$, $\delta_{\mathrm{d}}$) and \emph{probabilistic} model. The theoretical mean and standard deviation for $\lr\to\infty$ according to Eq.~\eqref{eq:deltainfty} are denoted $\mu_{\mathrm{p,\infty}}$ and $\delta_{\mathrm{p,\infty}}^{\delta_h}$, respectively.
    \label{fig:peaksTPBT}
    }
\end{figure*}

\subsection{Three point bending simulations \label{sec:TPBT}}
The simulated specimens are concrete beams with or without a~central notch loaded in three point bending. The dimensions of the specimen are: depth $D=150$\,mm, length $L=720$\,mm, span $S=600$\,mm and thickness $b=40$\,mm. The notched variant has a~central notch of length $a_0=75$\,mm (across half of its depth). Only the central part of the beam is represented by the discrete model (450$\times$120\,mm$^2$ in the unnotched and 150$\times$85\,mm$^2$ in the notched case); the surrounding material is modeled by linear elastic finite elements in order to reduce computational time. The loading is driven by prescribed displacement at the top midspan of the beam, and the step size is controlled by the crack mouth opening displacement (CMOD). For the notched case, the CMOD is simply measured as the opening across the notch, while in the unnotched case, the openings arising over several short intervals along the bottom central part of the span are measured and the maximum of these is considered to be the CMOD. The beam geometry is shown in Fig.~\ref{fig:scheme}ab.

All the simulations are stable, allowing the computation of specimen responses up to complete separation. However, once the crack is long enough and approaches the region discretized by elastic finite elements, the simulation must be terminated. We therefore terminate the calculation as soon as the loading force decreases to one third of the peak load.

From the large number of data generated, we decided to present load--deflection curves of (i) the deterministic model and (ii) probabilistic model using standard deviation $\delta_h=0.14$. Load-deflection curves of the probabilistic models with different variance are similar. Average responses and bands of width one standard deviation on each side obtained from 100 different simulations are shown in Fig.~\ref{fig:LD_curves}. All the average responses for the notched case are almost identical but the standard deviation of the response depends on the autocorrelation length. Increase in the autocorrelation length leads to increase in the standard deviation of the response. For $\lr\rightarrow 0$, the standard deviation is almost identical to what is computed by the deterministic model. In the unnotched case, the autocorrelation length affects also the average response. The probabilistic models yield in some cases significantly lower maximum forces than the deterministic models; see for example $\lr=25$\,mm.

Let us now plot only the peak loads calculated using both the deterministic and probabilistic models with all considered standard deviations $\delta_h$. Fig.~\ref{fig:peaksTPBT} presents the maximum load from the unnotched beams (on the left hand side) and notched beams (on the right hand side). These data are also listed in Tab.~\ref{tab:TPBT} in the~\ref{sec:tables} for convenience. 

Similar shapes of strength evolution dependent on the autocorrelation length were published in \citet[][Fig.~20]{SyrKor15} using a~continuous model of notched three point bent beams of variable size. 

\ref{sec:C} shows peak load averages and standard deviations obtained with models of unnotched bending with different macroscopic characteristic lengths and randomization techniques.

\begin{figure*}
	\centering
	\includegraphics[width=\textwidth]{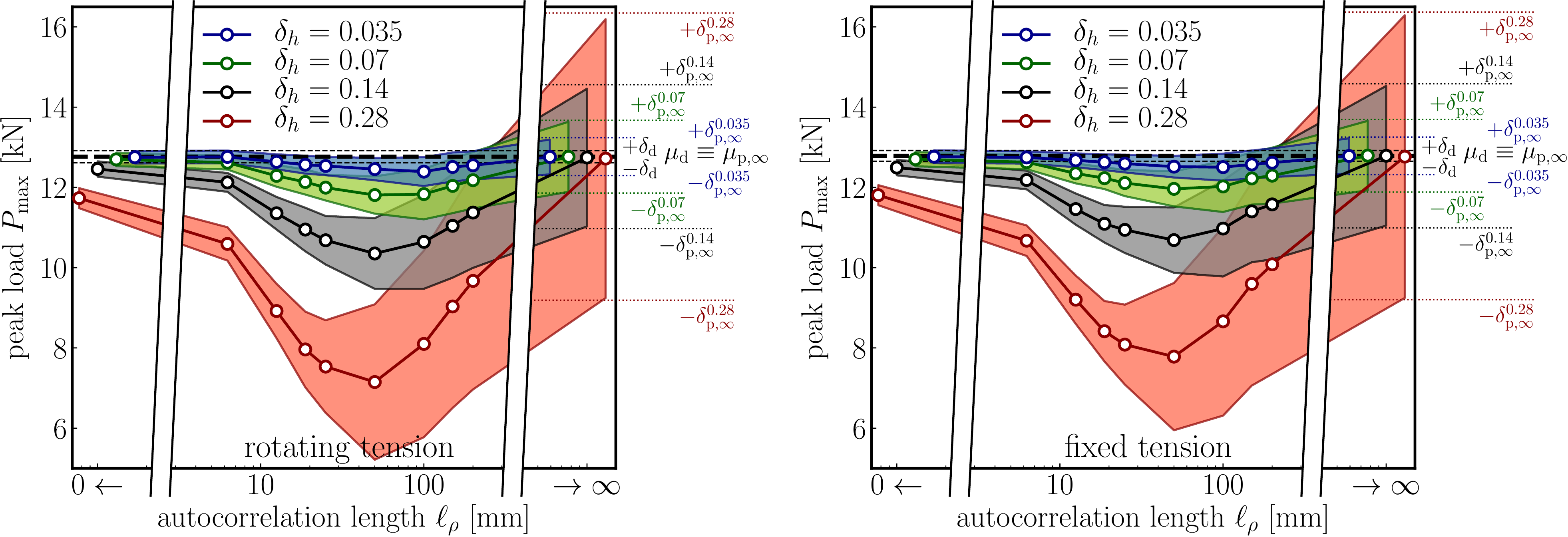}
	\caption{
		The mean value and standard deviation of the maximum load computed on a~virtual specimen loaded in uniaxial tension with rotating (left) and fixed (right) platens using the \emph{deterministic} model ($\mu_{\mathrm{d}}$, $\delta_{\mathrm{d}}$) and \emph{probabilistic} model. The theoretical mean and standard deviation for $\lr\to\infty$ according to Eq.~\eqref{eq:deltainfty} are denoted $\mu_{\mathrm{p,\infty}}$ and $\delta_{\mathrm{p,\infty}}^{\delta_h}$, respectively.
		\label{fig:peaksUT}
	}
\end{figure*}

\subsection{Uniaxial tension simulations}
The prisms loaded in uniaxial tension are depicted in Fig.~\ref{fig:scheme}cd. The dimensions are equal to the dimensions of the central part of three point bent unnotched specimens where the discrete model is used: length $S=450$\,mm, depth $D=120$\,mm and thickness $b=40$\,mm. Loading is realized via prescribed displacement. In case of the rigid platens, the degrees of freedom (DoFs) in the $x$ direction of all rigid bodies at the left and right specimen faces are directly prescribed, while the remaining DoFs are unconstrained (except for the restriction of the global rigid body motion). For rotating platens, the DoFs in the $x$ direction at the left and right specimen faces are constrained by prescribed platen displacement and free platen rotation, while the translational DoFs in the perpendicular directions and all rotational DoFs are again unconstrained.
The discrete model is applied throughout the whole volume, though the contacts at the left and right specimen faces (those involving rigid bodies with a~kinematic boundary condition or kinematic constraint) have purely elastic behavior. Similarly as with unnotched bending, the tensile simulations are controlled by the CMOD, which is taken as the maximum opening calculated over several virtual gauges placed along the beam's central axis.
 
The simulations of uniaxial tension are rather unstable, with a~strong snap-back. The probabilistic models with reasonable autocorrelation length and higher variances are typically more stable since the position of macrocrack localization is predetermined by the random field. In the remaining cases, the simulation typically reaches a~high number of iterations at the localization point as the position of macrocrack has a~number of equally suited candidates. The convergence criteria were often relaxed in order to proceed. Sometimes, the post-localization response could not be obtained with reasonable confidence and therefore the simulation was withdrawn from the results. In total, 100 deterministic and $10\times4\times100$ (10 autocorrelation lengths and 4 variances) probabilistic simulations were conducted, out of which 64 (1.6\%) were removed. The simulations of uniaxial tension were terminated soon after the peak load was reached (at 90\% of the peak load), since the postpeak response is less reliable.
 
The statistical characteristics of the maximum load from all the model variants are presented in Fig.~\ref{fig:peaksUT}. The data are again listed in Tab.~\ref{tab:UT} in the~\ref{sec:tables} for the sake of convenience. Note that (generally speaking) the beams with fixed platens are capable of transforming higher loads compared to those with rotating platens, which is in accordance with experimental evidence, e.g. \citep[sec. 4.1.3.2]{Mier96}. Additionally, we have also studied prisms with half depth ($D=60$\,mm), which is a~configuration with reduced 2D effects related to stress redistribution. Results from these simulations are listed in Tab.~\ref{tab:HUT} in the\ref{sec:tables}. The half-depth beams behave similarly to the full-depth beams and we will not comment on them hereinafter. They are, however, used in the companion paper, Part II~\citep{VorEli19II}.
 
\section{Discussion: load capacity \label{sec:peakload}}

Before commenting on selected results obtained for particular autocorrelation lengths $\lr$, we first present simple ideas about what governs the peak load. The sequence of cracks for an~increasing deflection depends on the positions and orientations of the contacts, and on their material parameters. These parameters also include the multipliers -- random variables that represent the discretized random field $h(\xx)$. The loading process  involves the redistribution of local forces among the surrounding contacts. Not one, but several contacts within the FPZ need to be at least partially damaged before a~macrocrack may propagate. All the contacts within the current FPZ are therefore involved in a~certain \emph{averaging} of their random capacities. The volume of averaging is the volume of the FPZ with a~size related to  the macroscopic characteristic length, $\ld$. Therefore, the weighted \emph{average} over the volume related to deterministic length is the governing variable determining the structural strength. The second fundamental mechanism determining structural strength is the \emph{weakest-link} that selects from the all possible positions of the FPZ the one with the worst combination of stress and strength.

When $\lr \to \infty$, the effect of the additional variability in probabilistic models is very easy to predict. Due to the selected method for the simultaneous randomization of local $\ft$ and $\Gt$ (Alternative II in Eq.~\eqref{eq:scaling2}), the damage evolution at individual contacts and therefore also the final crack patterns of the probabilistic models are strictly identical to those obtained with the deterministic models (for the same aggregate layouts). The forces at which these events occur are, however, linearly dependent on the value of random variable $h(\xx)=h$.  There are therefore two \emph{independent} sources of response variance in the probabilistic model: the variance inherent to the deterministic model and the variance due to random multiplier $h$. Due to this independence, one can simply calculate the mean and standard deviation of the peak load of the probabilistic model ($\mu_{\mathrm{p,\infty}}$ and $\delta_{\mathrm{p,\infty}}$) as the \emph{product} of two independent random variables. One is the random peak load in the deterministic model and the other is the strength multiplier $h$. The mean and standard deviation of the peak load for $\lr\rightarrow\infty$ read \citep{Goodman:variance:60}
\begin{align}
    \mu_{\mathrm{p,\infty}}
    &=
    \mu_{\mathrm{d}}  \mu_h
    =
    \mu_{\mathrm{d}} \\
    \delta_{\mathrm{p,\infty}}
    &=
    \sqrt{\delta_{\mathrm{d}}^2 \delta_h^2 + \mu_{\mathrm{d}}^2 \delta_h^2 + \delta_{\mathrm{d}}^2 \mu_h^2}
    =
    \sqrt{\delta_{\mathrm{d}}^2 \delta_h^2 + \mu_{\mathrm{d}}^2 \delta_h^2 + \delta_{\mathrm{d}}^2}
    \label{eq:deltainfty}
\end{align}
where we considered the selected mean of $h$: $\mu_h=1$. These analytically predicted values are shown in Figs.~\ref{fig:peaksTPBT} and \ref{fig:peaksUT} on the right hand side of each graph. The horizontal lines correspond relatively well to numerically computed values. The differences are attributed to the solver, which does not scale its nonlinear steps with multiplier $h$ and, in the case of uniaxial tension, to missing simulations deleted because of convergence issues (see the previous section). We conclude that when FPZ size is much lower compared to autocorrelation length ($\ld\ll\lr$), additional randomness has no effect on the evolution of the fracture process.

At the other extreme when the autocorrelation length $\lr \to 0$, the probabilistic models deliver a~mean value and standard deviation for the peak load that are almost identical to those obtained with the deterministic models. This is true for all models - notched and unnotched bending, as well as rotating and fixed tension. The unit-mean random field with basically symmetrical distribution seems to randomly modify the properties of the contacts which are scattered anyway due to their random orientations. This additional randomness averages out within an~FPZ of non-negligible size. We conclude that in the case of an~extremely short autocorrelation length with respect to the deterministic characteristic length ($\lr\ll\ld$),  randomness has theoretically no effect on the average strength or on its variance. The behavior of the probabilistic model tends towards that of the deterministic model. In our calculations, the finite size of discretization implies that the number of averaged independent variables is finite and not very high. Consequently, the averaging process does not remove randomness completely and a~weak effect of the weakest-link type exists: to some extent a~weaker spot may be found that leads to a~slight decrease in the mean value of the peak load compared to the deterministic model, see Fig.~\ref{fig:peaksTPBT} left. This effect is largely restricted in notched bent beams, where the presence of a~strong stress concentrator limits the sampling of possible FPZ locations.

The peak load for autocorrelation lengths lying in between zero and infinity is governed by the conflict between the weakest-link and averaging effect. The weakest-link searches for the worst combination of stress and strength. Shorter autocorrelation length implies more strength fluctuations, and thus there is a~higher probability of finding weaker regions. On the other hand, the effect of averaging increases with decreasing $\lr$ and thus ``fights'' against the weakest-link effect. There is clearly a~critical value of $\lr$ that produces large enough fluctuations and sufficiently mild averaging for the minimum structural strength to be attained. Let us now comment on the four different boundary conditions simulated.

The main difference is between the beam with the stress concentrator (notched bent beam) and the remaining beams. The notch suppresses the weakest-link effect, with the lowest average load capacity being 94.6\% of the deterministic strength achieved for $\delta_h=0.28$ and an~$\lr$ of about 6.25$\sim$12.5\,mm, see Tab.~\ref{tab:TPBT}. It is actually surprising that the weakest-link effect is active since the crack always starts from the notch. Nevertheless, there is apparently still some freedom in the choice of the crack direction and FPZ shape (see the discussion in Section~\ref{sec:crack_hist}). In the unnotched bent beam, the weakest-link domain is restricted to the area around the bottom midspan. The large freedom in FPZ position produces a~significantly lower minimum average load capacity: 69.3\% of the deterministic strength for $\delta_h=0.28$. The pronounced  weakest-link effect also increases the critical autocorrelation length (18.75$\sim$25\,mm, see Tab.~\ref{tab:TPBT}). Finally, the tensile loading ensures the largest domain for the weakest-link, yet the fixed platens constitutes an~additional constraint. Indeed, the minimum average load capacity corresponds to it; for $\delta_h=0.28$ it is 56.0\% or 60.9\% of the deterministic peak load for rotating or fixed platens, respectively. The maximized weakest-link effect gives the largest critical autocorrelation length (50$\sim$100\,mm, see Tab.~\ref{tab:UT}).

\section{Size and shape of the fracture process zone
    \label{sec:ActiveZone}
}

\begin{figure}[tb!]
\centering
\includegraphics[width=\textwidth]{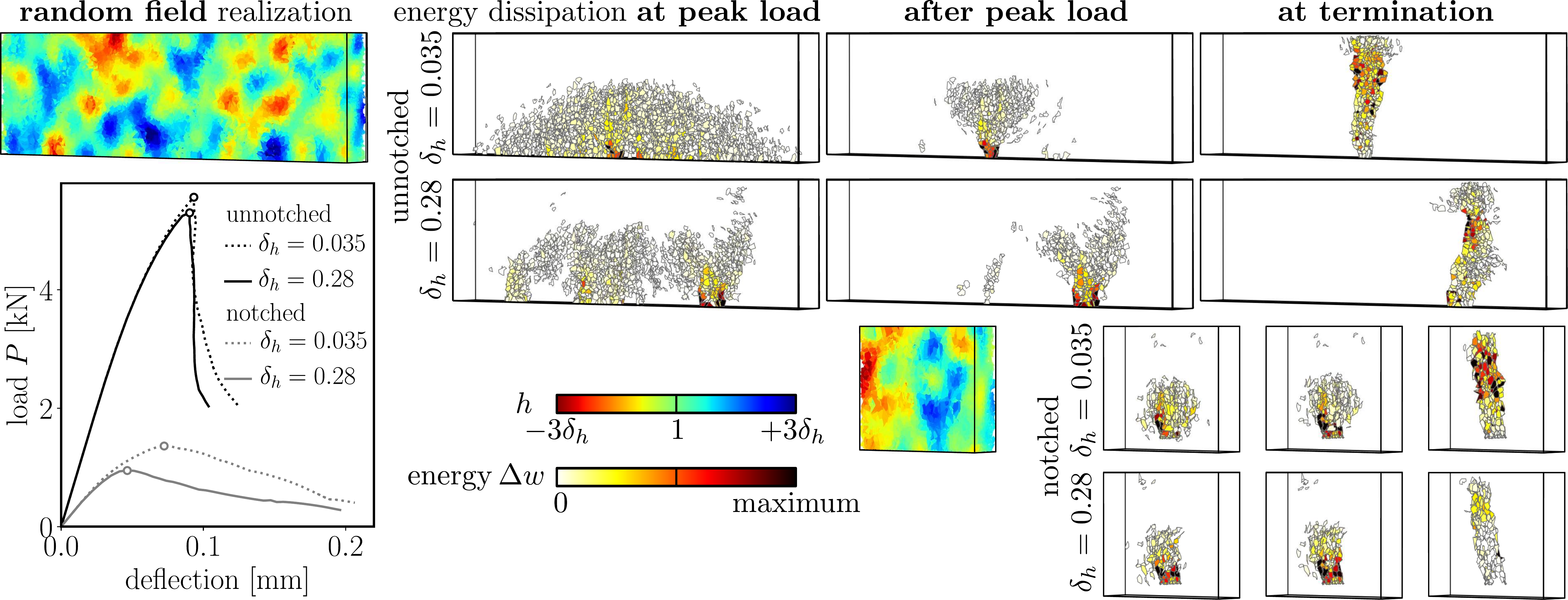}
\caption{Distribution of energy dissipation within a~single computational step at the inter-particle contacts of one realization of the \emph{probabilistic} model ($\lr=25\,$mm) of a~bent beam with and without a~notch for two random field standard deviations ($\delta_h\in\left\{0.035,\,0.28\right\}$).  \label{fig:TPBT_cracks}}
\end{figure}

\begin{figure}[tb!]
\centering
\includegraphics[width=\textwidth]{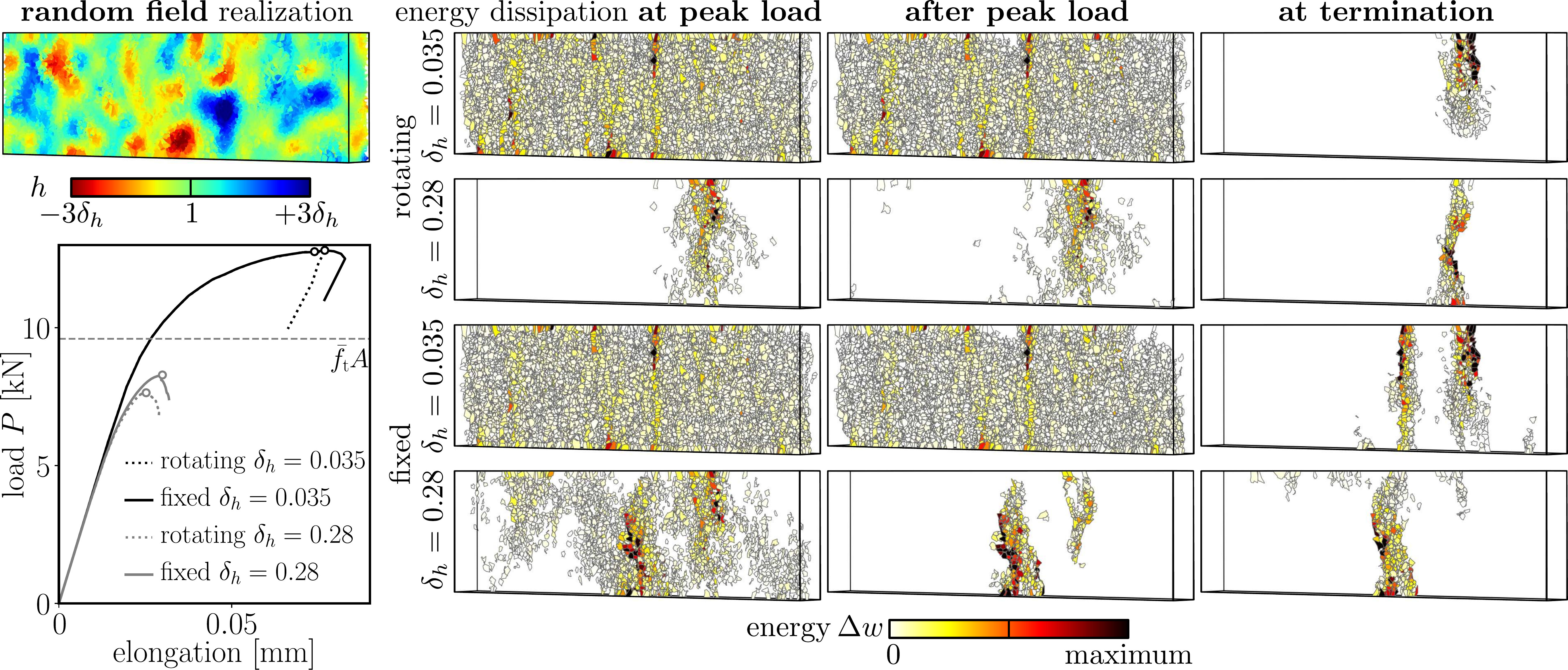}
\caption{Distribution of energy dissipation within a~single computational step at the inter-particle contacts of one realization of the \emph{probabilistic} model ($\lr=25\,$mm) of rotating and fixed tension for two random field standard deviations ($\delta_h\in\left\{0.035,\,0.28\right\}$). \label{fig:UT_cracks}}
\end{figure}

The fundamental role of the FPZ in the fracture process in quasibrittle materials is now widely accepted. There have been numerous attempts to measure the size and shape of the FPZ experimentally. To name just a~few techniques, excellent results are obtained by Moiré interferometry \citep{CedPol-83,CedPol-87}, digital image correlation \citep{SkaSyk-11,LinYua-14,BhoRay19} and acoustic emission \citep{MihNom96,OtsDat00,MurPra-10,RenLiu-20} techniques, or their combination \citep{AlaLou-15,GuoAla-17,LiFan-20}. There are also numerical models that attempt to describe the FPZ \citep{GraJir10,FraVes-13,BenPou-16,TarMak-17} with results strongly resembling the experimental measurements.

\begin{figure*}[!tb]
\centering
\includegraphics[width=\textwidth]{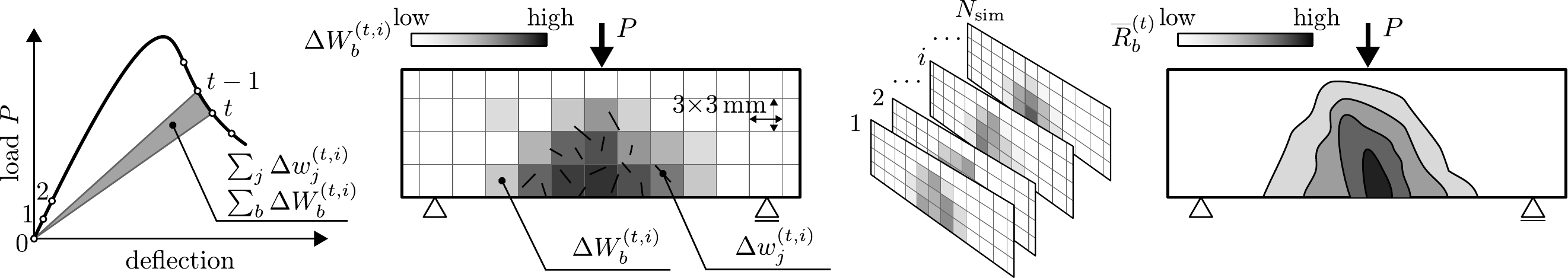}
\caption{Illustration of the total energy dissipated in the $i$th realization at the $t$th loading step (left) and a~summation of the energies dissipated  at individual contacts over bins, $\Delta w\rightarrow \Delta W\rightarrow\overline{R}$ (right).    \label{fig:AZ}
}
\end{figure*}

The mesoscale model at hand provides a~detailed description of fracture processes. With the help of such a~calculation, we shall now proceed with an~evaluation of the effect of the autocorrelation length on the size and shape of the nonlinear zone, and thus the deterministic length $\ld$ as well. We analyze the microcracking zone by looking at energy dissipation in loading step $t$ (see Fig.~\ref{fig:AZ}). The volume that dissipates energy during this loading step is considered to be the \emph{fracture process zone} at time $t$.

The energy dissipated at particular contact with contact area $A$ and length $l$ at step $t$ (from time $t-1$ to $t$) is difference between total dissipated energies up to times $t$ and $t-1$
\begin{align}
\Delta w^{(t)} = w_{\text{dis}}(t)-w_{\text{dis}}(t-1)
\end{align}
The cumulative energy dissipation is calculated as the difference between the total input energy and stored strain energy which, for our model based on damage mechanics, becomes
\begin{align}
w_{\text{dis}}(t) 
= \underbrace{Al\int\limits_0^t \bm{s}^T(t')\dot{\bm{e}}(t')\dd{t'}}_{\text{total energy}} - \underbrace{\frac{Al}{2}\bm{s}^T(t) \bm{e}(t)}_{\text{strain energy}}
\approx \frac{Al}{2}\left[\sum_{o=1}^t (\bm{s}^{(o)} + \bm{s}^{(o-1)})^T  (\bm{e}^{(o)} - \bm{e}^{(o-1)}) - \bm{s}_{(t)}^T \bm{e}^{(t)}\right]
\end{align}
where the continuous integration over time is replaced by summation over time steps $o$. Energies dissipated in a~simulation $i$ within a~chosen step $t$ of the nonlinear solver at a~particular contact $j$ with a~centroid $\bm{c}_j$ in discrete models are denoted $\Delta w_j^{(t,i)}$.

These energy differences are plotted in Fig.~\ref{fig:TPBT_cracks} for specimens loaded in three point bending; only the discrete models' domains are shown. One simulation number ($i=1$), a~single autocorrelation length ($\lr=3/160\,$m) and two standard deviations ($\delta_h\in\left\{0.035,\,0.28\right\}$) are chosen for demonstration purposes. The figure shows the $i$th random field realization, which is the same for both variances when normalized by $\delta_h$. The load-deflection curves from individual simulations are plotted in the graph on the left. On the right hand side energies $\Delta w_j^{(t,i)}$ are depicted in a~3D view for 3 steps -- the one in which the maximum load is attained ({\bf peak load}), the very next one ({\bf after peak load}) and the step in which the simulation was terminated ({\bf at termination}).

The random field realization was purposely selected so it does not have a~weak spot in the midspan of the unnotched beam. As a~consequence, the crack location is moved by a~large distance towards the right hand side support for $\delta_h=0.28$ but remains in the midspan for $\delta_h=0.035$. The peak force reached by these two models with different $\delta_h$ is similar, as for $\delta_h=0.28$ it is triggered by weaker material but in a~region of lower stress. If the weak spot appears at the midspan, the crack starts there for both $\delta_h$ and a~significant difference in peak loads is obtained because a~midspan stress region of different strength is decisive. The energy plots of the unnotched beam also show the following  typical pattern: in deterministic model or probabilistic models with low $\delta_h$ a~large zone of distributed cracking develops before the peak load is reached. After this, the macrocrack immediately localizes, resulting in a~steep drop in the loading force, and possibly also in snap-back. In probabilistic models with higher $\delta_h$, one can see several cracks that have already localized before the peak at the position of weak spots. One of them later grows while the others close. Also, the drop in the loading force after the load capacity has been reached is not that severe. In notched beams, not much difference can be seen between high and low $\delta_h$. It seems that for high $\delta_h$, a~slightly lower number of contacts dissipate energy and the crack is more localized. From the viewpoint of the analytical model developed in the companion paper, Part II \citep{VorEli19II}, it is important that for both notched and unnotched bending only one macrocrack is active at the peak load. 

Similar observations have been made for prisms loaded in uniaxial tension.  Fig.~\ref{fig:UT_cracks} shows results obtained for simulation number $i=0$. For low $\delta_h$, the cases with both rotating and fixed platens produce a~large zone of distributed cracking that suddenly localizes in macrocracks. Surprisingly, this localization often does not appear at the maximum load; it appears later. The beams with fixed platens that restrict rotation usually produce multiple macrocracks. When higher $\delta_h$ is applied, the distributed damage zone is replaced by gradual localization into one (rotating platens) or several (fixed platens) cracks that grow or close as the simulation proceeds. 

The observations made with a~single simulation are interesting but solid knowledge can only be gained from a~statistical evaluation of the result. We will now describe the procedure of processing the spatial distribution of energy dissipation in all $\Ns=100$ simulations. Due to the relatively small specimen thickness, all the energies dissipated at the contacts in three-dimensional space are projected (lumped) onto the $x$-$y$ plane. We divide this plane into $N_b$ bins of size $3\times3$~mm$^2$ and sum all the contact energies $\Delta w_j^{(t,i)}$ that spatially pertain to these bins. The total energy dissipated in bin $b$ occupying the region $\left[\xx_b\right]$ is then
\begin{align}
    \Delta  W_b^{(t,i)}
    =
    \sum\limits_{j:\bm{c}_j \in \left[\xx_b\right]} \Delta  w_j^{(t,i)}
\end{align}
Such an~approach has been applied to every single simulation $i \in \{1,\ldots,\Ns\}$. However, in order to obtain statistically relevant data, we attempt to sum the energies across all $\Ns$ simulations. Several problems appear in such a~summation:
\begin{itemize}
\item The step length is adaptive implying that the step numbers are not synchronized across the simulations $i \in \{1,\ldots,\Ns\}$. The summation is therefore performed not using the same step numbers, but using steps at the same level of relative loading force with respect to the maximum load, $P_{\max}^{(i)}$. The variable $t$ is hereinafter understood to represent the step within which the loading force reached the value $tP_{\max}^{(i)}$ either in the prepeak or postpeak stage.

\begin{figure}[tb]
\centering
\includegraphics[width=\textwidth]{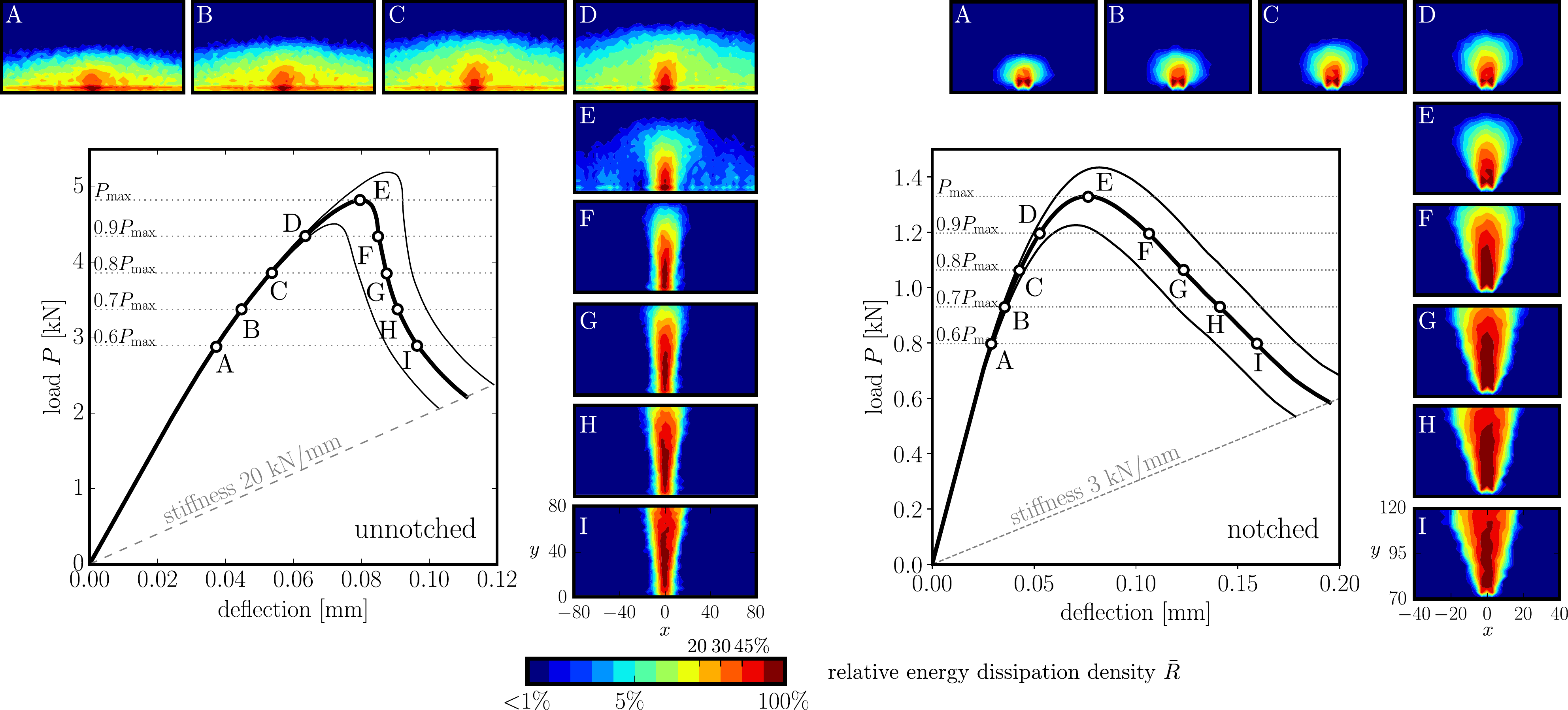}
\caption{Average load-deflection curves and $\pm$ standard deviations for notched and unnotched bent beams ($\lr=12.5$\,mm, $\delta_h=0.14$), as well as relative energy dissipation densities visualizing FPZs at different stages of the fracture process.
\label{fig:contours_evolution}}
\end{figure}

\item The variability in the step length of individual simulations makes the contributions to the energy summation incomparable. The remedy applied is to normalize the energy contribution by total dissipated energy in a~chosen step, and therefore each simulation contributes by the unit amount in total. Instead of summing $\Delta W_b^{(t,i)}$ directly, the relative values $\Delta \overline{W}_b^{(t,i)}$ are used.
\begin{align}
    \Delta \overline{W}_b^{(t,i)} = \frac{\Delta  W_b^{(t,i)}}{\sum\limits_{r=1}^{N_{b}} \Delta W_r^{(t,i)}}
\end{align}

\item If no stress concentrator is present, the location of the FPZ is random. For unnotched specimens loaded in bending, we have performed horizontal shifts of the coordinate system of individual simulations to obtain overlapping FPZs centered at the location of the macrocracks. The macrocrack locations are detected for each simulation at the last simulation step. The $x$ coordinate of a~centroid of dissipated energies $\Delta w_j^{(\mathrm{final},i)}$ in the bottom-most layer of depth 25\,mm is found and taken as a~central point for shifting the energy contribution across simulations. In the case of notched beams, no such shifting is necessary as the FPZ has a~predetermined position at the notch tip.  The summed dissipation densities in the bins are $R_b^{(t)} = \sum_{j=1}^{N_{\mathrm{sim}}}\Delta \overline{W}_b^{(t,j)}$.

For tensile specimens, the crack can finally localize anywhere in the volume. Some shifting (in two directions) combined with flipping is probably possible, but the results seem to be very sensitive to the centralization procedure. We keep the statistical results limited only to the bent specimen as the application of the summation process to the tensile simulations is questionable.

\item
    Finally, after summing all the contributions from all \Ns=100 simulations for each type
    of bent specimen, the results are normalized for the second time. All the  $R_b^{(t)}$ values
    are divided by their maximum
\begin{align}
\overline{R}_b^{(t)} = \frac{R_b^{(t)}}{\max\limits_{r} R_r^{(t)}}
\end{align}
The original data in the bins did not have any relevant unit, so no information is lost by such normalization. The advantage is that the operation makes the data comparable and easy to display. The maximum value of 1 will be referred to as 100\%. The process of energy summation in space and across simulations is schematically shown in Fig.~\ref{fig:AZ}.
\end{itemize}

Fig.~\ref{fig:contours_evolution} presents the evolution of the FPZ for both notched and unnotched bending at different loading levels $t$. The FPZs are computed for $\lr=25$\,mm and related to the average load-deflection curves.  The red color marks areas with the highest relative energy dissipation density, $\overline{R}_b$, while the dark blue color shows bins with less than 1\%  compared to the maximum.  In the unnotched case, one can see the initial widely distributed energy dissipation, its localization at the peak (stage ``E''), and its propagation in localized form in the postpeak. The figures are generally similar to those reported from experiments with acoustic emission. The width of the localized FPZ is similar in the notched and unnotched case (note that the snapshots have different length units).

\begin{figure}[tb]
	\centering
	\includegraphics[width=15cm]{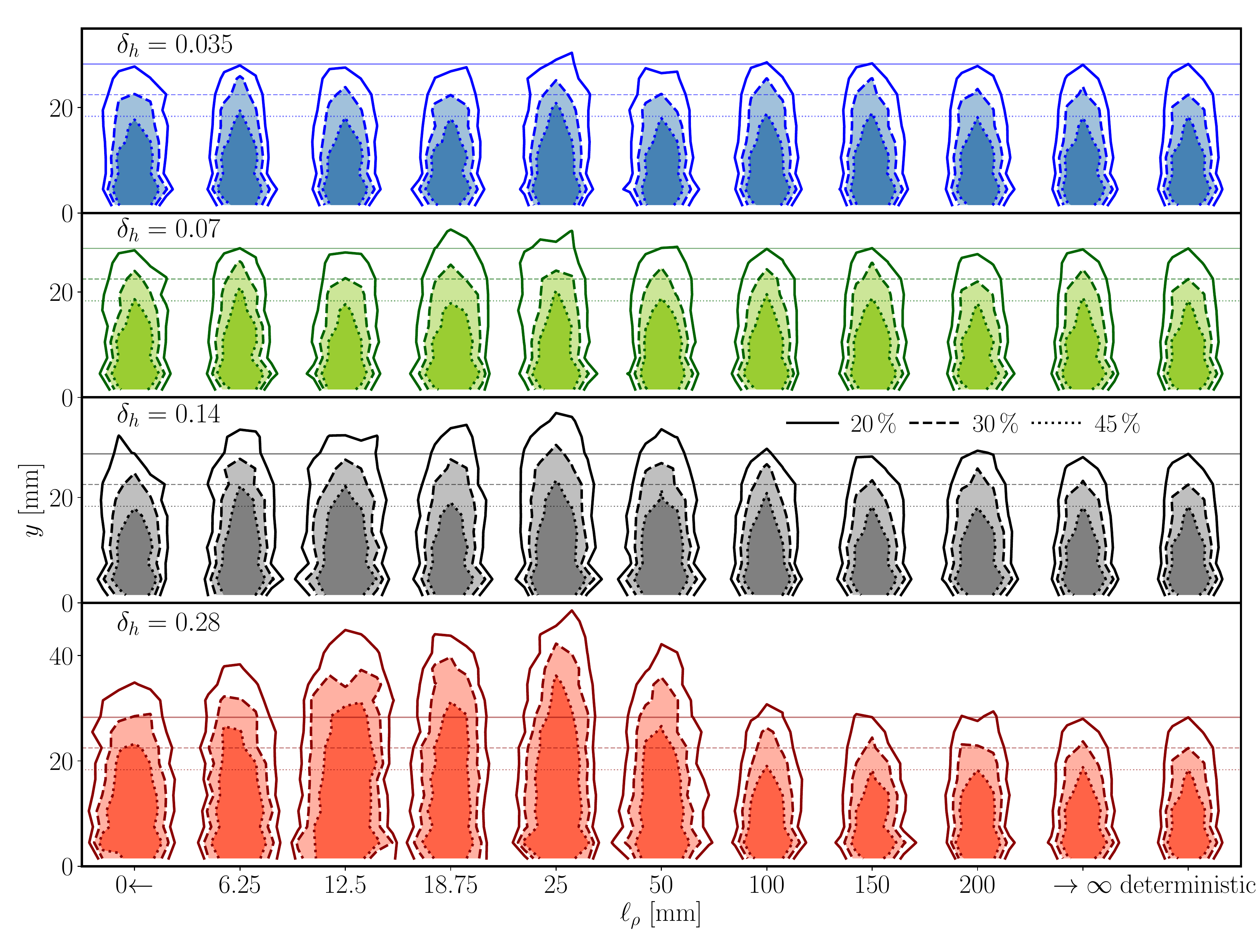}
	\caption{The averaged FPZ at the peak load for models of \emph{unnotched} bent beams with different autocorrelation lengths, $\lr$, and standard deviations of the random field, $\delta_h$. The rightmost shape and the horizontal lines in the graphs belong to the deterministic simulations. 
		\label{fig:contours_unnotched}}
\end{figure}

\begin{figure}[tb]
	\centering
	\includegraphics[width=15cm]{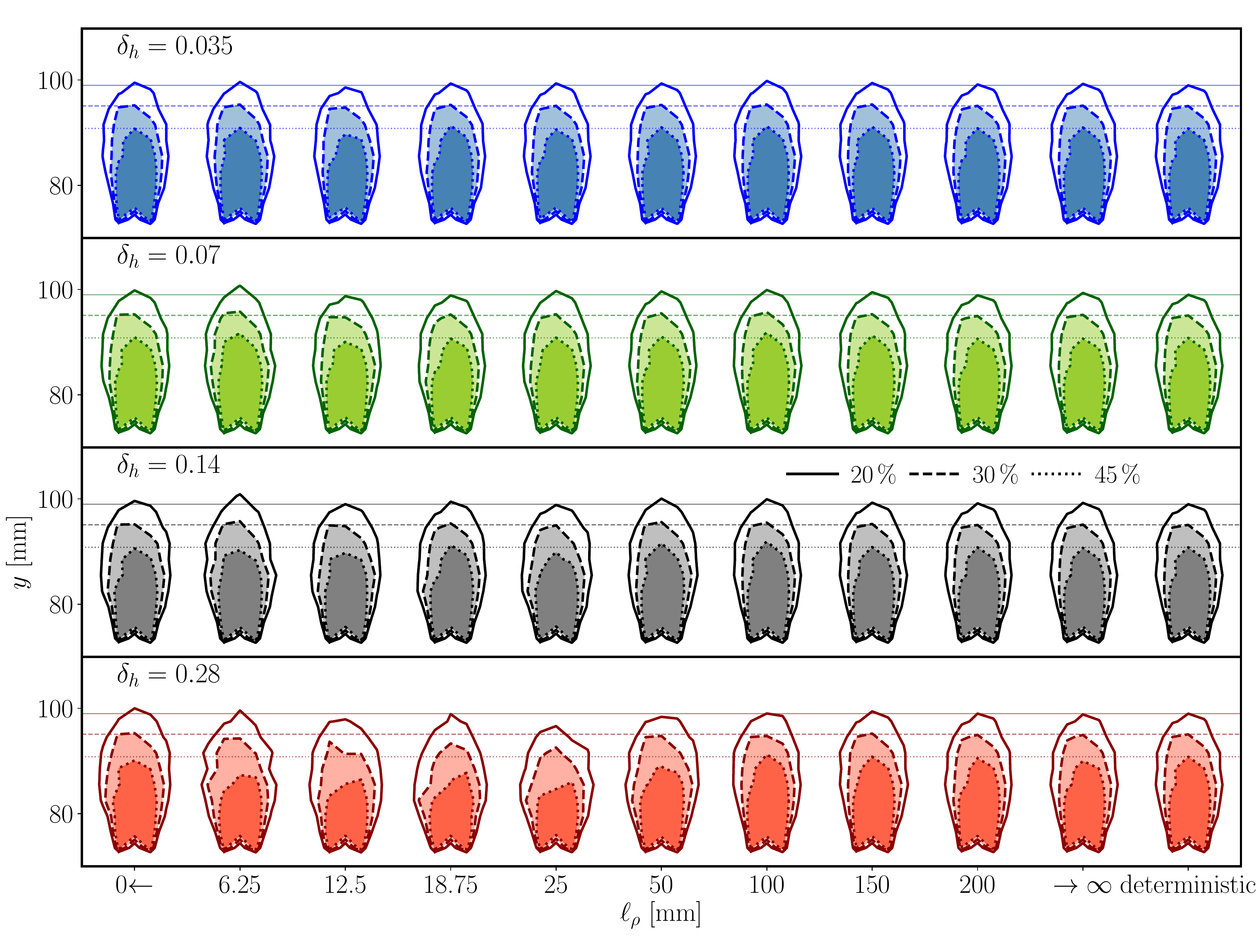}
	\caption{The averaged FPZ at the peak load for models of \emph{notched} bent beams with different autocorrelation lengths, $\lr$, and standard deviations of the random field, $\delta_h$. The rightmost shape and the horizontal lines in the graphs belong to the deterministic simulations. 
		\label{fig:contours_notched}}
\end{figure}

Of particular interest is the FPZ at the peak load as it can identify the material volume responsible for the magnitude of the peak load where the local averaging takes place. The FPZ may allow the identification of the region and the number of contacts that simultaneously (in parallel) contribute to energy dissipation. Such a~region is indeed used in the companion paper, Part II \citep{VorEli19II}, for the derivation of an~analytical model capable of  delivering statistics regarding peak load. We would like to observe the effect of the autocorrelation length and random field variance on the FPZ. The average FPZs at the peak load are computed for different autocorrelation lengths and variances, as well as for the deterministic model. The contours of volumes containing 20, 30 and 45\% of the maximum relative energy dissipation density ($\max R_r^{(\mathrm{peak})}=1$) are shown in Figs.~\ref{fig:contours_unnotched} and \ref{fig:contours_notched}. 

Though the graphics are not very smooth, indicating that improvement is possible by adding even more simulations, there are clearly some trends emerging. The figures show that the notched case is neither influenced by the autocorrelation length nor the variance of the random field. On the contrary, in the unnotched configuration the size of the FPZ exhibits a~mild dependence on both $\lr$ and $\delta_h$. For the extreme choice $\lr = \infty$, the FPZ size and shape remain identical to that produced by the deterministic model thanks to the constant mesoscopic characteristic length preserved in Alternative II. In another extreme, $\lr = 0$, when the contacts obtain an~additional independent random multiplier, the averaging within the FPZ is effective enough to average out almost all the spatial variability and the FPZ size and shape again tend to resemble those from the deterministic model. However, the FPZ size tends to increase when the autocorrelation length is around its critical value. This effect is pronounced for greater variances of $h$.

The increase in FPZ depth  is attributed to the vertical position of the weak material region. For sufficiently high $\delta_h$, the weak spot appearing above the bottom surface may be decisive and trigger localization. In such a~case, the dissipating zone is extended from the weak spot down to the bottom surface, and therefore its depth increases.

There is also a~mild increase in width for higher $\delta_h$ around the critical $\lr$. We expected the opposite effect -- as the gradient of the random field increases with $\delta_h$, the FPZ should shrink into the deep valleys of low strength. This phenomenon is indeed seen in individual simulations such as those shown in Figs.~\ref{fig:TPBT_cracks} and \ref{fig:UT_cracks}. On the other hand, the FPZ in individual simulations becomes more tortuous and inclined. Additionally, when averaging over many of these tortuous FPZ is conducted, the contours of energy dissipation actually widen. The aggregated energy dissipation may create the wrong impression that the FPZ becomes wider with increasing $\delta_h$, when it actually becomes more localized in individual simulations.

The FPZ is in general dictated by (i) the aggregate size distribution  (mostly by $d_{\max}$), (ii) the constitutive relation and its parameters, including their possible random spatial fluctuation, and (iii) the boundary conditions. The only difference between unnotched and notched bending (and so between Figs.~\ref{fig:contours_unnotched} and \ref{fig:contours_notched}) is in the boundary conditions. Considering the large difference caused by the presence or absence of a~notch, there is a~surprising similarity in the FPZ size and shape for the deterministic models (Figs.~\ref{fig:contours_unnotched} and \ref{fig:contours_notched} can be directly compared as they are plotted in the same scale). The bottom wedge in the FPZ from notched simulations is an~artefact of the semi-regular placement of aggregates in the vicinity of the notch tip. 

\ref{sec:C} discusses and shows FPZs obtained with models with different macroscopic characteristic lengths and randomization techniques.

\section{Macrocrack locations in bending simulations \label{sec:crack_hist}}

\begin{figure}[tb!]
\centering
\includegraphics[width=\textwidth]{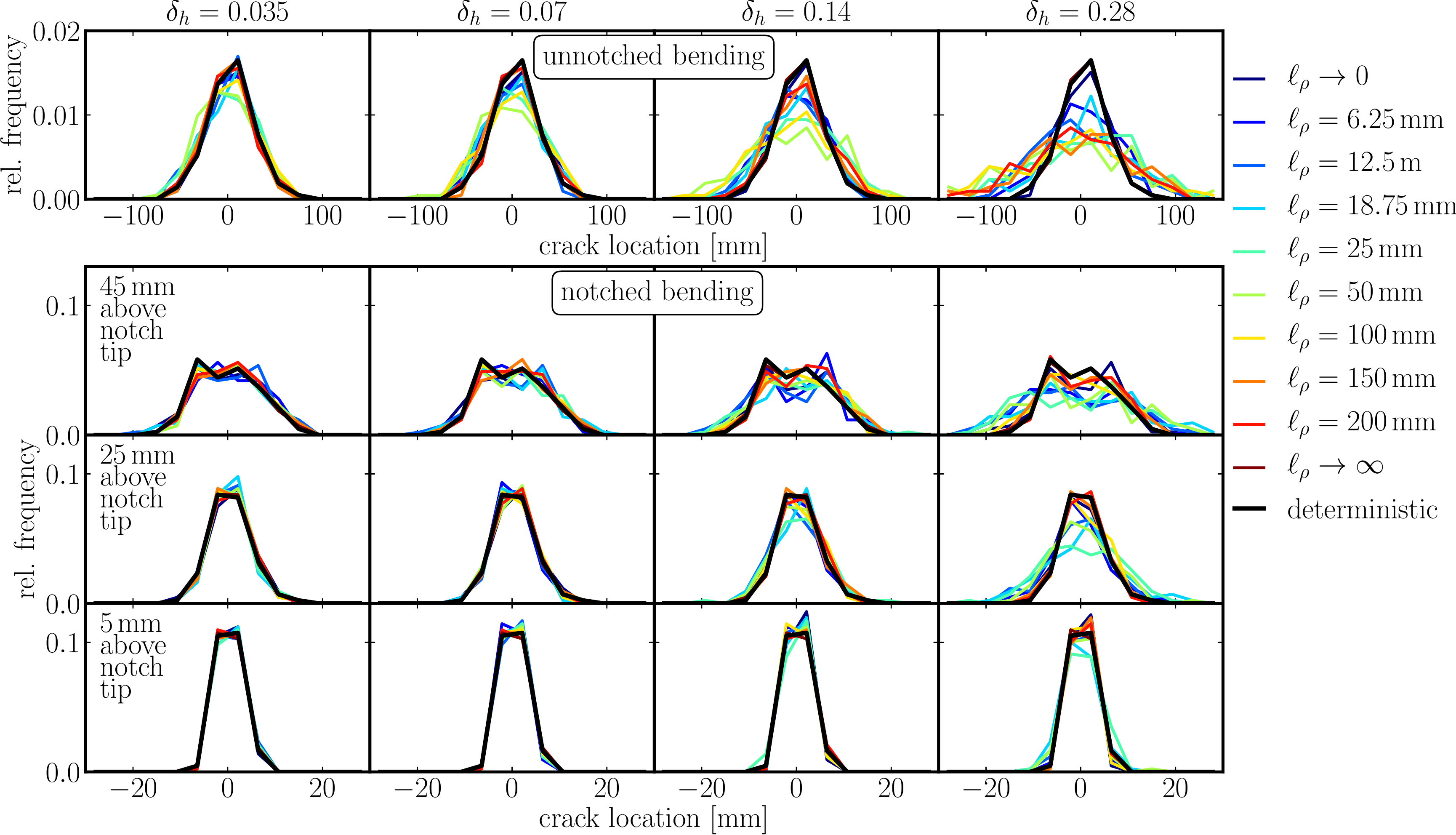}
\caption{Histograms of macrocrack locations along the beam span (location 0 is the midspan) for each autocorrelation length as well as the deterministic model. Top: unnotched bending measured 5\,mm above the bottom surface; bottom: notched bending measured 5, 25 and 45\,mm above the notch tip. \label{fig:cracks_hist}}
\end{figure}

The analytical model proposed in the companion paper Part II \citep{VorEli19II} requires approximate knowledge of the size of the domain in which the macrocrack shall be expected. This section attempts to deliver such information based on the results from the discrete mesoscale model. The macrocrack location is identified by the centroid of the instantaneous energy dissipation densities $\Delta w^i_j$ in the terminal step of the simulation. The centroid is computed in strips of 10\,mm in depth over the full specimen width and span. 

Histograms of the locations along the span from all the bending simulations are plotted in Fig.~\ref{fig:cracks_hist}. The top row of the figure shows crack location in the bottom-most strip of the unnotched beams. The probabilistic models have cracks located in wider interval compared to the deterministic model, but larger differences are seen only for high random field variance. The short autocorrelation lengths only provide modest widening of the crack location interval, because (i) the randomness is partly averaged out inside the FPZ, and (ii) the high random field fluctuation provides a~high density of weak material spots and the crack initiates in one of them close to the midspan, where there is also a~high stress level.  The widest interval is found for an~autocorrelation length of about 50$\sim$100\,mm, where the averaging effect is weak and strength fluctuations produce only a~few weak material regions that might be quite far from the midspan. The even larger autocorrelation lengths produce a~random field with low gradients that cannot create a~region of sufficiently weak material far from the midspan to initiate the macrocrak, and thus the crack location interval shrinks. 

In notched bending simulations, the crack is forced to propagate from the notch tip. However, the histograms of macrocrack location in strips 5, 25 and 45\,mm above the notch tip (the three bottom rows in Fig.~\ref{fig:cracks_hist}) get wider with distance from the tip and show that there is still some freedom in the crack inclination. This allows the crack to bypass the strong material regions and to run through the weak ones. The same effects as described for unnotched bending (averaging and the weakest-link effect) are also active in the notched bending model. The widest histograms of crack locations are obtained for autocorrelation lengths of around $\lr=18.75$\,mm. This length is shorter than in the case of the unnotched bending because the stress profile gradients are greater and thus the weakest-link effect is suppressed.

The crack locations in tensile simulations were not processed. The visual comparison suggest they are created uniformly throughout the whole volume. The attempt to confirm this by centroid calculation is spoiled by (i) the frequent appearance of several macrocraks in one simulation, (ii) early termination of the calculation prior to the development of a~large macrocrack, and also by (iii) the initiation of the macrocrack at different strips over the specimen depth.

\section{Conclusions}

A~probabilistic mesoscale discrete model is employed to simulate the fracture of concrete specimens loaded in three point bending (with and without a~notch) and in uniaxial tension (with fixed and rotating loading platens). The primary focus is on the interaction between the autocorrelation length and the variance of the random field determining spatial fluctuations of material parameters on the one hand, and the characteristic length of the fracture process arising from concrete mesostructure on the other. The statistical evaluation of the simulation campaign is presented in terms of the peak loads and fracture process zones that dissipate energy (experience inelastic processes) at the peak load.

In three point bending simulations with a~notch, the average peak load is found to be insensitive to the spatial variability in material parameters. A~mild sensitivity only appears in the case of large random field variance.  The reason is that the stress concentration is so severe that the crack is forced to propagate from one specific location (the notch tip) and the ability of the spatial material variability to change its location is largely suppressed. However, the standard deviation of the peak load changes with the autocorrelation length due to averaging of the fluctuations within the FPZ. The size and shape of the FPZ is independent of the applied autocorrelation length or variance of the random field.

In unnotched three point bending simulations, the average peak loads in probabilistic models with both short and long autocorrelation lengths (with respect to characteristic length) are approximately equal to the average maximum load obtained with the deterministic model. For intermediate autocorrelation lengths, the average strengths of the probabilistic model are lower than the average strength of the deterministic model. This drop is pronounced with higher variance of the random field. The minimum of the mean strength occurs when the autocorrelation length is close to the material's characteristic length. Such an~autocorrelation length enables the structure to sample the position of the fracture process zone into the weakest region, but averaging within the fracture process zone is not yet severe enough to filter the randomness out. The standard deviation follows the same trend as in the notched case. The FPZ length increases for greater random field variance and autocorrelation lengths close to the characteristic length, as the decisive weak region may occur further away from the bottom surface.

The uniaxial tensile simulation behaves similarly to the unnotched bending version: the weakest-link effect is even more pronounced, so the downward trend in the load capacity gets emphasized. The model behaves in agreement with experimental observations, which show that the fixed loading platens leads to a~slightly higher load capacity compared to the rotating platens. It also typically produces multiple cracks, while the rotating platens leads to a~single one. Due to the large variability in crack position and shape, the statistical processing of the FPZ is abandoned for the uniaxial tension.

The companion paper Part II \citep{VorEli19II} introduces relatively simple yet robust analytical model capable of capturing the presented results both qualitatively and quantitatively.

\section*{ACKNOWLEDGEMENT}
The authors acknowledge financial support provided by the Czech Science Foundation under projects Nos. GA19-12197S and GC19-06684J.

\appendix
\section{Different macroscopic characteristic length and scaling of $G_{\mathrm{t}}$ \label{sec:C}}

\begin{figure*}[tb!]
	\centering
	\includegraphics[width=14cm]{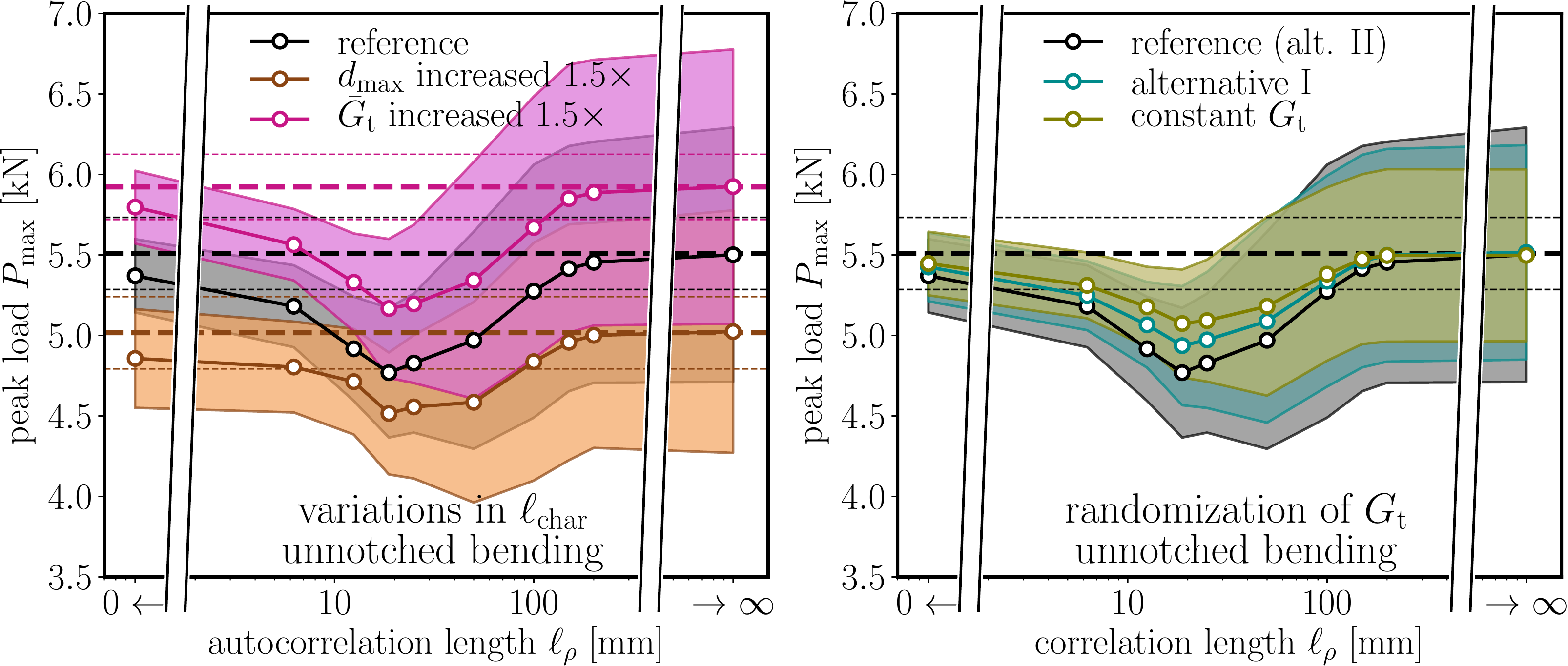}
	\caption{
		The mean value and standard deviation of the maximum load computed on unnotched beams loaded in three point bending using the \emph{deterministic} (horizontal dashed lines) and \emph{probabilistic} model with $\delta_h=0.14$. The reference model variant refers to the results from subsection~\ref{sec:TPBT}. It is compared to four other model variants differing in the macroscopic characteristic length and a~randomization procedure.
		\label{fig:peaksC}
	}
\end{figure*}

The choice of the randomization of the model was not based on any experimental evidence. One can easily imagine that the mesoscopic fracture energy and tensile strength are independent or correlated in a~different ways. One can also study the effect of randomness on material with different macroscopic characteristic length. In order to include at least some of these situations into our analysis, we decided to run four studies with model where
\begin{itemize}
\item the maximum aggregate diameter was increased 1.5 times to become $d_{\max}=15$\,mm
\item the mesoscopic fracture energy was increased 1.5 times to become 37.5\,J/m$^2$
\item the randomization of the model was performed according to alternative I (Eq.~\ref{eq:scaling1})
\item the fracture energy in the probabilistic model was left constant, equal to $\oGt$
\end{itemize}
The first two cases modify the macroscopic characteristic length of the material, while the remaining two cases vary the correlation between $\ft$ and $\Gt$. 

\begin{figure}[tb]
	\centering
	\includegraphics[width=15cm]{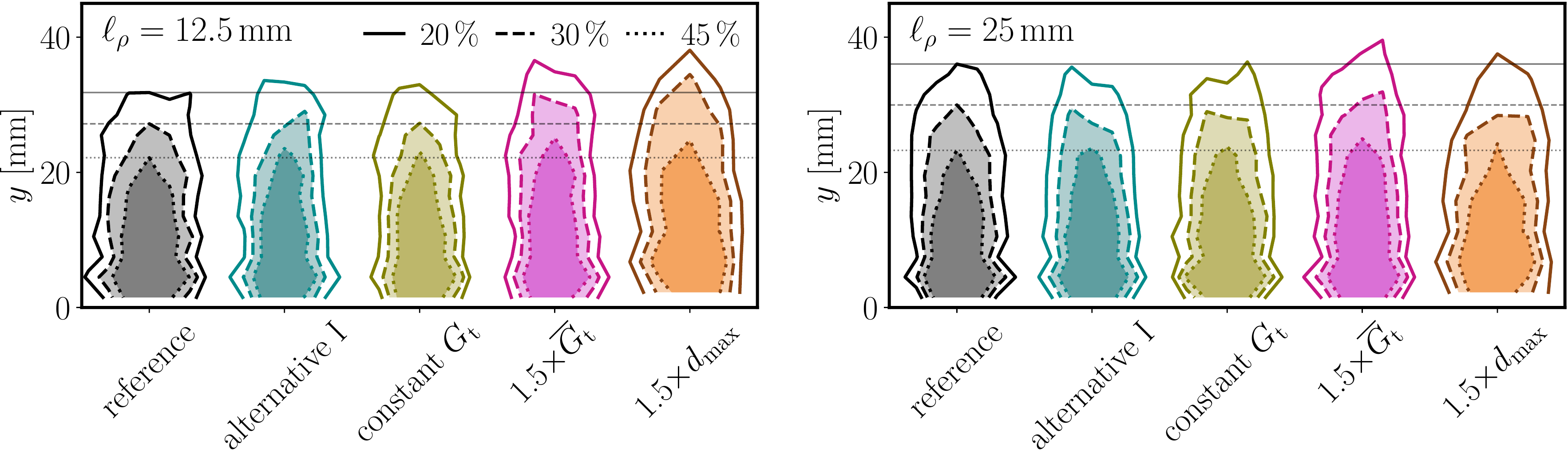}
	\caption{The averaged FPZ at the peak load for models of \emph{unnotched} bent beams with
		(i) different randomization procedure (alternative I), (ii) constant $\Gt$ during randomization, (iii) increased mean value of mesoscopic fracture energy $\oGt$ and (iv) increased maximum aggregate size $d_{\max}$. The reference case is the model with randomization alternative II  presented in Fig.~\ref{fig:contours_unnotched}.
		$\delta_h=0.14$	in all the cases.
		\label{fig:contours_C}}
\end{figure}

Only simulations with beams loaded in unnotched bending with $\delta_h=0.14$ were analyzed. Averages and standard deviations of the maximum load are plotted in Fig.~\ref{fig:peaksC}. The results are qualitatively the same as what was observed previously.

Models with increased $\Gt$ exhibit higher peak load thanks to direct enhancement of mesoscale material properties. Increasing maximum aggregate diameter results in decrease of maximum loads because the crack branching and tortuosity is reduced; the main factor is probably reduction in the number of aggregates in the $z$ direction (thickness).  Also behavior of the remaining cases, constant $\Gt$ and randomization according to alternative I, can be easily explained. If the weakest-link effect is activated, contacts with random parameter $h$ bellow one are predominantly damaged. Therefore, the alternative II with squared $h$ multiplier of $\oGt$ dissipates less energy than alternative I with linear multiplier $h$. The largest energy dissipation occurs for constant $\Gt$. The maximum loads from these three models are in this order for intermediate autocorrelation lengths where weakest-link effect takes place.

Fig.~\ref{fig:contours_C} includes selected contours of the energy dissipation. It shows FPZ at the peak load for autocorrelation lengths $\lr=12.5$\,mm (the lowest peak load) and $\lr=25$\,mm (the deepest FPZ for alternative II). Both models with different randomization strategy (alternative I and constant $\Gt$) yields more or less identical size and shape of the FPZ. The model with increased $\oGt$ seems to have slightly deeper FPZ as a~consequence of higher cohesive forces acting at the damaged contacts. Also increase of $d_{\max}$ seems to slightly increase depth of the FPZ. However, the results in the last case of increased $d_{\max}$ might be corrupted by the averaging procedure that is for this particular case performed on 1.5$\times$ enlarged bins in an~attempt to obtained comparable results.

\section{The average values and standard deviations of peak loads \label{sec:tables}}

\begin{table}
\centering
\caption{Mean value and standard deviation estimations of peak loads for probabilistic models of bent unnotched and notched beams. All the values are in percentages [\%] of the peak load mean value achieved by the deterministic model, which is $\mu_{\mathrm{d}}$=5.509\,kN for unnotched and $\mu_{\mathrm{d}}$=1.356\,kN for notched beams. The grey cells indicate the critical load capacity  for each variance. \label{tab:TPBT}}
\vspace{2mm}
\tabcolsep 4.5pt
\begin{tabular}{r|rr|rr|rr|rr|rr|rr|rr|rr}
\Xhline{4\arrayrulewidth}
&\multicolumn{8}{c|}{unnotched bending}&\multicolumn{8}{c}{notched bending}\\\hline\hline
\Tstrut deter.&\multicolumn{8}{c|}{$\mu_{\mathrm{d}}$=100\,\%\,(5.509\,kN)\quad $\delta_{\mathrm{d}}$=4.06}&\multicolumn{8}{c}{$\mu_{\mathrm{d}}$=100\,\%\,(1.356\,kN)\quad $\delta_{\mathrm{d}}$=3.93}\\\hline\hline
\Tstrut prob.&\multicolumn{2}{c|}{$\delta_h=0.035$}&\multicolumn{2}{c|}{$\delta_h=0.07$}&\multicolumn{2}{c|}{$\delta_h=0.14$}&\multicolumn{2}{c|}{$\delta_h=0.28$}&\multicolumn{2}{c|}{$\delta_h=0.035$}&\multicolumn{2}{c|}{$\delta_h=0.07$}&\multicolumn{2}{c|}{$\delta_h=0.14$}&\multicolumn{2}{c}{$\delta_h=0.28$}\\
$\lr$ [m]&$\mu_{\mathrm{p}}$&$\delta_{\mathrm{p}}$&$\mu_{\mathrm{p}}$&$\delta_{\mathrm{p}}$&$\mu_{\mathrm{p}}$&$\delta_{\mathrm{p}}$&$\mu_{\mathrm{p}}$&$\delta_{\mathrm{p}}$&$\mu_{\mathrm{p}}$&$\delta_{\mathrm{p}}$&$\mu_{\mathrm{p}}$&$\delta_{\mathrm{p}}$&$\mu_{\mathrm{p}}$&$\delta_{\mathrm{p}}$&$\mu_{\mathrm{p}}$&$\delta_{\mathrm{p}}$\Bstrut\\\hline
\Tstrut$0\leftarrow$	&	99.7	&	4.0	&	99.1	&	3.9	&	97.5	&	4.1	&	92.7	&	4.4	&	100.1	&	4.1	&	99.9	&	4.1	&	99.4	&	4.1	&	97.6	&	4.8	\\
1/160	&	99.4	&	4.0	&	98.1	&	3.9	&	94.1	&	4.6	&	83.9	&	5.4	&	\cellcolor{gray}99.9	&	4.2	&	99.8	&	4.7	&	98.7	&	5.9	&	\cellcolor{gray}94.6	&	8.0	\\
2/160	&	98.8	&	4.4	&	96.3	&	4.8	&	89.3	&	5.9	&	74.6	&	8.2	&	100.0	&	4.5	&	\cellcolor{gray}99.7	&	5.5	&	\cellcolor{gray}98.6	&	7.8	&	95.2	&	12.8	\\
3/160	&	98.6	&	4.9	&	\cellcolor{gray}95.0	&	5.7	&	\cellcolor{gray}86.6	&	7.3	&	70.6	&	10.7	&	100.0	&	4.5	&	99.8	&	6.1	&	99.0	&	9.8	&	96.0	&	17.7	\\
1/40	&	\cellcolor{gray}99.0	&	4.3	&	96.1	&	5.4	&	87.6	&	7.8	&	\cellcolor{gray}69.3	&	11.8	&	100.0	&	5.0	&	99.8	&	7.0	&	99.0	&	11.4	&	95.8	&	19.8	\\
1/20	&	99.2	&	4.5	&	97.1	&	6.6	&	90.2	&	12.2	&	72.3	&	21.4	&	100.0	&	4.8	&	100.0	&	7.2	&	99.8	&	12.9	&	98.8	&	25.5	\\
0.10	&	99.5	&	5.4	&	98.8	&	8.0	&	95.8	&	14.3	&	84.1	&	26.9	&	100.0	&	5.3	&	100.0	&	7.9	&	100.0	&	14.0	&	99.8	&	27.0	\\
0.15	&	99.7	&	4.8	&	99.2	&	7.4	&	98.3	&	13.8	&	93.7	&	27.0	&	100.0	&	5.1	&	100.0	&	7.7	&	100.1	&	14.0	&	100.0	&	27.3	\\
0.20	&	99.7	&	5.1	&	99.7	&	7.5	&	99.0	&	13.6	&	95.9	&	26.8	&	100.0	&	5.2	&	100.0	&	7.6	&	100.0	&	13.4	&	100.0	&	25.6	\\
$\rightarrow\infty$	&	99.9	&	5.3	&	99.9	&	8.0	&	99.9	&	14.4	&	99.8	&	27.8	&	100.0	&	5.2	&	100.0	&	8.0	&	100.0	&	14.3	&	100.1	&	27.8	\\\Xhline{4\arrayrulewidth}
\end{tabular}
\end{table}

\begin{table}
\centering
\caption{Mean value and standard deviation estimations of peak loads for probabilistic models of uniaxial tension in \emph{ prisms with a~depth of 120\,mm} with rotating and fixed platens. All the values are in percentages [\%] of the peak load mean value achieved by the deterministic model, which is $\mu_{\mathrm{d}}$=12.77\,kN for rotating and $\mu_{\mathrm{d}}$=12.79\,kN for fixed platens. The grey cells indicate the critical load capacity  for each variance. \label{tab:UT}}
\vspace{2mm}
\tabcolsep 4.5pt
\begin{tabular}{r|rr|rr|rr|rr|rr|rr|rr|rr}
\Xhline{4\arrayrulewidth}
&\multicolumn{8}{c|}{rotating tension, depth=120\,mm}&\multicolumn{8}{c}{fixed tension, depth=120\,mm}\\\hline\hline
\Tstrut deter.&\multicolumn{8}{c|}{$\mu_{\mathrm{d}}$=100\,\%\,(12.77\,kN)\quad $\delta_{\mathrm{d}}$=1.19}&\multicolumn{8}{c}{$\mu_{\mathrm{d}}$=100\,\%\,(12.79\,kN)\quad $\delta_{\mathrm{d}}$=1.05}\\\hline\hline
\Tstrut prob.&\multicolumn{2}{c|}{$\delta_h=0.035$}&\multicolumn{2}{c|}{$\delta_h=0.07$}&\multicolumn{2}{c|}{$\delta_h=0.14$}&\multicolumn{2}{c|}{$\delta_h=0.28$}&\multicolumn{2}{c|}{$\delta_h=0.035$}&\multicolumn{2}{c|}{$\delta_h=0.07$}&\multicolumn{2}{c|}{$\delta_h=0.14$}&\multicolumn{2}{c}{$\delta_h=0.28$}\\
$\lr$ [m]&$\mu_{\mathrm{p}}$&$\delta_{\mathrm{p}}$&$\mu_{\mathrm{p}}$&$\delta_{\mathrm{p}}$&$\mu_{\mathrm{p}}$&$\delta_{\mathrm{p}}$&$\mu_{\mathrm{p}}$&$\delta_{\mathrm{p}}$&$\mu_{\mathrm{p}}$&$\delta_{\mathrm{p}}$&$\mu_{\mathrm{p}}$&$\delta_{\mathrm{p}}$&$\mu_{\mathrm{p}}$&$\delta_{\mathrm{p}}$&$\mu_{\mathrm{p}}$&$\delta_{\mathrm{p}}$\Bstrut\\\hline
\Tstrut$0\leftarrow$	&	99.9	&	1.2	&	99.5	&	1.3	&	97.6	&	1.5	&	91.9	&	2.0	&	99.8	&	1.2	&	99.3	&	1.4	&	97.7	&	1.5	&	92.3	&	2.0	\\
1/160	&	99.9	&	1.2	&	98.8	&	1.3	&	95.0	&	1.8	&	83.0	&	3.3	&	99.7	&	1.2	&	98.8	&	1.4	&	95.3	&	1.8	&	83.5	&	3.0	\\
2/160	&	99.0	&	1.5	&	96.3	&	2.1	&	89.0	&	3.2	&	69.9	&	5.3	&	99.1	&	1.4	&	96.6	&	1.9	&	89.6	&	3.0	&	72.0	&	4.7	\\
3/160	&	98.4	&	1.7	&	95.1	&	2.5	&	85.8	&	4.3	&	62.4	&	7.4	&	98.7	&	1.5	&	95.6	&	2.1	&	86.8	&	3.7	&	65.8	&	5.9	\\
1/40	&	98.2	&	1.8	&	94.0	&	2.5	&	83.7	&	4.7	&	59.0	&	9.0	&	98.5	&	1.7	&	94.7	&	2.6	&	85.6	&	4.5	&	63.2	&	7.8	\\
1/20	&	97.6	&	2.2	&	92.5	&	3.6	&	\cellcolor{gray}81.1	&	6.9	&	\cellcolor{gray}56.0	&	15.2	&	97.9	&	2.1	&	\cellcolor{gray}93.6	&	3.4	&	\cellcolor{gray}83.6	&	6.4	&	\cellcolor{gray}60.9	&	14.4	\\
0.10	&	\cellcolor{gray}97.1	&	2.8	&	\cellcolor{gray}92.7	&	4.9	&	83.4	&	9.2	&	63.4	&	18.2	&	\cellcolor{gray}97.8	&	2.6	&	94.0	&	5.0	&	85.8	&	9.3	&	67.7	&	18.4	\\
0.15	&	98.0	&	2.8	&	94.3	&	5.3	&	86.5	&	10.1	&	70.8	&	19.8	&	98.3	&	2.7	&	95.6	&	5.1	&	89.2	&	9.9	&	75.0	&	19.8	\\
0.20	&	98.3	&	2.8	&	95.4	&	5.5	&	89.1	&	10.8	&	75.8	&	21.2	&	98.7	&	2.8	&	96.1	&	5.5	&	90.5	&	10.7	&	78.9	&	21.4	\\
$\rightarrow\infty$	&	99.9	&	3.4	&	100.0	&	6.8	&	99.8	&	13.4	&	99.6	&	27.2	&	100.0	&	3.4	&	100.0	&	6.8	&	100.0	&	13.6	&	99.8	&	27.5	\\
\Xhline{4\arrayrulewidth}
\end{tabular}
\end{table}

\begin{table}
\centering
\caption{Mean value and standard deviation estimations of peak loads for probabilistic models of uniaxial tension in \emph{specimens with a~depth of 60\,mm} with rotating and fixed platens. All the values are in percentages [\%] of the peak load mean value achieved by the deterministic model, which is $\mu_{\mathrm{d}}$=6.26\,kN for rotating and $\mu_{\mathrm{d}}$=6.28\,kN for fixed platens. The grey cells indicate the critical load capacity  for each variance. \label{tab:HUT}}
\vspace{2mm}
\tabcolsep 4.5pt
\begin{tabular}{r|rr|rr|rr|rr|rr|rr|rr|rr}
\Xhline{4\arrayrulewidth}
&\multicolumn{8}{c|}{rotating tension, depth=60\,mm}&\multicolumn{8}{c}{fixed tension, depth=60\,mm}\\\hline\hline
\Tstrut deter.&\multicolumn{8}{c|}{$\mu_{\mathrm{d}}$=100\,\%\,(6.26\,kN)\quad $\delta_{\mathrm{d}}$=1.81}&\multicolumn{8}{c}{$\mu_{\mathrm{d}}$=100\,\%\,(6.28\,kN)\quad $\delta_{\mathrm{d}}$=1.54}\\\hline\hline
\Tstrut prob.&\multicolumn{2}{c|}{$\delta_h=0.035$}&\multicolumn{2}{c|}{$\delta_h=0.07$}&\multicolumn{2}{c|}{$\delta_h=0.14$}&\multicolumn{2}{c|}{$\delta_h=0.28$}&\multicolumn{2}{c|}{$\delta_h=0.035$}&\multicolumn{2}{c|}{$\delta_h=0.07$}&\multicolumn{2}{c|}{$\delta_h=0.14$}&\multicolumn{2}{c}{$\delta_h=0.28$}\\
$\lr$ [m]&$\mu_{\mathrm{p}}$&$\delta_{\mathrm{p}}$&$\mu_{\mathrm{p}}$&$\delta_{\mathrm{p}}$&$\mu_{\mathrm{p}}$&$\delta_{\mathrm{p}}$&$\mu_{\mathrm{p}}$&$\delta_{\mathrm{p}}$&$\mu_{\mathrm{p}}$&$\delta_{\mathrm{p}}$&$\mu_{\mathrm{p}}$&$\delta_{\mathrm{p}}$&$\mu_{\mathrm{p}}$&$\delta_{\mathrm{p}}$&$\mu_{\mathrm{p}}$&$\delta_{\mathrm{p}}$\Bstrut\\\hline
\Tstrut $0\leftarrow$	& 99.7 &	1.9 &	99.0 &	1.9 &	96.9 &	2.2 &	90.1 &	2.5 &	99.8 &	1.5 &	99.2 &	1.6 &	97.3 &	1.9 &	91.1 &	1.9\\
1/160 & 99.4 &	1.9 &	98.2 &	2.3 &	93.5 &	2.8 &	79.8 &	4.2 &	99.6 &	1.6 &	98.6 &	1.9 &	94.1 &	2.6 &	80.9 &	3.5\\
2/160 & 98.6 &	1.9 &	96.0 &	2.2 &	87.6 &	3.8 &	66.0 &	6.8 &	98.9 &	1.7 &	96.3 &	2.2 &	88.9 &	3.8 &	68.5 &	5.7\\
3/160 & 98.1 &	2.1 &	94.3 &	3.2 &	83.6 &	5.2 &	58.2 &	9.0 &	98.6 &	2.1 &	94.9 &	3.0 &	85.2 &	5.0 &	62.1 &	8.4\\
1/40 & 97.6 &	2.5 &	92.6 &	3.7 &	80.8 &	6.2 &	\cellcolor{gray}54.0 &	11.8 &	98.1 &	2.1 &	93.5 &	3.6 &	82.7 &	5.8 &	\cellcolor{gray}58.3 &	11.0\\
1/20 & \cellcolor{gray}96.9 &	2.6 &	\cellcolor{gray}91.6 &	4.4 &	\cellcolor{gray}79.8 &	8.1 &	55.2 &	16.1 &	\cellcolor{gray}97.2 &	2.7 &	\cellcolor{gray}92.7 &	4.4 &	\cellcolor{gray}82.3 &	8.0 &	59.3 &	16.2\\
0.10 & 97.0 &	3.0 &	92.9 &	5.2 &	83.9 &	9.9 &	64.6 &	19.3 &	97.6 &	3.2 &	93.6 &	5.3 &	85.5 &	10.0 &	67.8 &	19.5\\
0.15 & 98.1 &	2.6 &	95.2 &	4.8 &	87.9 &	9.6 &	73.0 &	19.8 &	98.2 &	2.7 &	95.2 &	4.8 &	89.3 &	9.8 &	75.9 &	19.9\\
0.20 & 98.4 &	3.2 &	95.7 &	5.1 &	89.5 &	10.1 &	77.1 &	21.1 &	98.7 &	3.0 &	96.3 &	5.2 &	90.8 &	10.2 &	78.9 &	20.6\\
$\rightarrow\infty$ & 99.9 &	4.0 &	99.9 &	7.2 &	99.9 &	14.0 &	99.5 &	27.6 &	100.0 &	3.9 &	100.0 &	7.1 &	100.1 &	14.0 &	100.0 &	27.6\\

\Xhline{4\arrayrulewidth}
\end{tabular}
\end{table}

\end{document}